\documentclass{aa}
\usepackage{graphicx}
\begin{document}

\title{Lithium and H$\alpha$ in stars and brown dwarfs of
  $\sigma$\,Orionis \thanks{Based on observations made with the
    following telescopes: 3.5-m telescope at the Spanish-German Calar
    Alto Observatory (Spain) operated by the Max-Planck-Institut f\"ur
    Astronomie in Heidelberg (Germany); 2.5-m Isaac Newton telescope
    operated on the island of La Palma by the Isaac Newton Group in
    the Spanish Observatorio del Roque de Los Muchachos of the
    Instituto de Astrof\'\i sica de Canarias; 2.1-m Otto Struve
    telescope at McDonald Observatory (U.S.A.); and the 10-m Keck II
    telescope of the W.\,M. Keck Observatory, which is operated as a
    scientific partnership among the California Institute of
    Technology, the University of California and the National
    Aeronautics and Space Administration (the Observatory was made
    possible by the generous financial support of the W.\,M. Keck
    Foundation).  }  }

\subtitle{}

\author{M.\,R. Zapatero Osorio,
       \inst{1,2,3}
        V.\,J.\,S. B\'ejar,
       \inst{4}
        Ya. Pavlenko,
       \inst{5}
        R. Rebolo,
       \inst{4,6}
        C. Allende Prieto,
       \inst{7}
        E.\,L. Mart\'\i n
       \inst{8}
       \and
        R.\,J. Garc\'\i a L\'opez
       \inst{4,9}
       }

\offprints{M.\,R. Zapatero Osorio ~ \email{mosorio@gps.caltech.edu, mosorio@laeff.esa.es}}

\institute{Division of Geological and Planetary Sciences, MS 150-21, 
   California Institute of Technology, Pasadena, CA 91125, U.S.A.\\
  \and
  Mount Wilson Observatory, 740 Holladay Road, Pasadena, CA 91106, U.S.A.\\
  \and
  Currently at: LAEFF-INTA, ESA Satellite Tracking Station, PO 50727, 
  E-28080 Madrid, Spain\\
  \and
  Instituto de Astrof\'\i sica de Canarias, E-38200 La Laguna, Tenerife, 
  Spain\\
  \and
  Main Astronomical Observatory of Academy of Sciences of Ukraine, 
  Golosiiv woods, Kyiv-127, 03680, Ukraine\\
  \and
  Consejo Superior de Investigaciones Cient\'\i ficas, Madrid, Spain\\
  \and
  McDonald Observatory and Department of Astronomy, University of 
  Texas, Austin, TX 78712-1083, U.S.A.\\
  \and
  Institute for Astronomy, Univ. of Hawaii at Manoa, Honolulu, 
  HI 96822, U.S.A.\\
  \and Departamento de Astrof\'\i sica, Universidad de La Laguna,
  E-38206 La Laguna, Tenerife, Spain }

\date{Received [date];  Accepted [date]}

\abstract{We present intermediate- and low-resolution optical spectra
  around H$\alpha$ and Li\,{\sc i} $\lambda$6708\,\AA~for a sample of
  25 low mass stars and 2 brown dwarfs with confirmed membership in
  the pre-main sequence stellar $\sigma$\,Orionis cluster. Our
  observations are intended to investigate the age of the cluster. The
  spectral types derived for our target sample are found to be in the
  range K6--M8.5, which corresponds to a mass interval of roughly
  1.2--0.02\,$M_{\odot}$ on the basis of state-of-the-art evolutionary
  models. Radial velocities (except for one object) are found to be
  consistent with membership in the Orion complex. All cluster members
  show considerable H$\alpha$ emission and the Li\,{\sc i} resonance
  doublet in absorption, which is typical of very young ages. We find
  that our pseudo-equivalent widths of H$\alpha$ and Li\,{\sc i}
  (measured relative to the observed local pseudo-continuum formed by
  molecular absorptions) appear rather dispersed (and intense in the
  case of H$\alpha$) for objects cooler than M3.5 spectral class,
  occurring at the approximate mass where low mass stars are expected
  to become fully convective. The least massive brown dwarf in our
  sample, S\,Ori\,45 (M8.5, $\sim$0.02\,$M_{\odot}$), displays
  variable H$\alpha$ emission and a radial velocity that differs from
  the cluster mean velocity. Tentative detection of forbidden lines in
  emission indicates that this brown dwarf may be accreting mass from
  a surrounding disk. We also present recent computations of Li\,{\sc
    i} $\lambda$6708\,\AA~curves of growth for low gravities and for
  the temperature interval (about 4000--2600\,K) of our sample. The
  comparison of our observations to these computations allows us to
  infer that no lithium depletion has yet taken place in
  $\sigma$\,Orionis, and that the observed pseudo-equivalent widths
  are consistent with a cluster initial lithium abundance close to the
  cosmic value. Hence, the upper limit to the $\sigma$\,Orionis
  cluster age can be set at 8\,Myr, with a most likely value around
  2--4\,Myr.   }

\titlerunning{Lithium and H$\alpha$ in low mass cluster members of
  $\sigma$\,Orionis} 
\authorrunning{Zapatero Osorio et al.}
\maketitle

%

\section{Introduction}
Deep photometric and spectroscopic searches in various nearby
star-forming regions and young open clusters have revealed populations
of very low mass stars ($\le$0.3\,$M_{\odot}$), brown dwarfs (see
Basri \cite{basri00} for a review) with masses below the
hydrogen-burning mass limit ($\sim$0.075\,$M_{\odot}$) and
planetary-mass objects (Najita, Tiede, \& Carr \cite{najita00}; Lucas
et al$.$ \cite{lucas01}; Zapatero Osorio et al$.$ \cite{osorio00})
smaller than the deuterium-burning threshold at 0.013\,$M_{\odot}$
(Saumon et al$.$ \cite{saumon96}). Age is one of the most relevant
parameters for their study and characterization. Traditionally, mass
estimates of cluster members rely on evolutionary models, which
predict luminosities and effective temperatures (hereafter $T_{\rm
  eff}$) as a function of time and mass. This procedure, namely
isochrone fitting to observational photometric diagrams, is highly age
dependent for the smallest objects because they coold down very
quickly (e.g., D'Antona \& Mazzitelli \cite{dantona94}; Burrows et
al$.$ \cite{burrows97}; Chabrier et al$.$ \cite{chabrier00}). To
better constrain masses, it is desirable to date clusters with higher
accuracy.

Lithium absorption features in optical spectra can be used as a tracer
of the stellar internal structure. In addition, lithium is valuable to
assess the age of stars in clusters. Dating young clusters and field
objects based on lithium analysis is a procedure nearly independent of
distance and reddening, in marked contrast to the isochrone-fitting
technique. Moreover, lithium isochrones do not significantly depend on
metallicity (D'Antona \cite{dantona00}), rendering the lithium dating
technique very powerful. Pre-main sequence stars with large convective
regions burn lithium efficiently on short time scales (see
Pinsonneault \cite{pin97} for a review) as soon as the temperature at
the base of the convective zone becomes hot enough to undergo the
nuclear reaction $^7$Li\,+\,p\,$\rightarrow$\,$^4$He\,+\,$\alpha$. Stars
smaller than the Sun require only about 10--15\,Myr to deplete this
element by one order of magnitude, and all M-type stars are observed
to have destroyed their lithium at ages around 20--40\,Myr (e.g.,
Pinsonneault, Kawaler \& Demarque \cite{pin90}; D'Antona \& Mazzitelli
\cite{dantona94}, \cite{dantona97}; Baraffe et al$.$
\cite{baraffe98}). Furthermore, lithium detections in fully convective
objects near and below the substellar limit have been successfully
used to constrain the ages of clusters like the Pleiades (Basri, Marcy
\& Graham \cite{basri96}; Mart\'\i n et al$.$ \cite{martin98a};
Stauffer, Schultz, \& Kirkpatrick \cite{stauffer98}), $\alpha$\,Persei
(Stauffer et al$.$ \cite{stauffer99}), and IC\,2391 (Barrado y
Navascu\'es et al$.$ \cite{barrado99}). Lithium dating, which is
fundamentally a nuclear age calibrator, can be considered reliable
even though some uncertainties (rotation, activity, mixing processes)
may affect theoretical calculations.

Recently, various groups have investigated the stellar and substellar
populations around the bright, massive and multiple O9.5V-type star
$\sigma$\,Orionis, which gives its name to the association (Walter et
al$.$ \cite{walter94}; Wolk \cite{wolk96}; Walter, Wolk, \& Sherry
\cite{walter98}; B\'ejar, Zapatero Osorio, \& Rebolo \cite{bejar99};
Zapatero Osorio et al$.$ \cite{osorio99}, \cite{osorio00}; B\'ejar et
al$.$ \cite{bejar01a}). These authors have adopted a cluster age
between 1\,Myr and 7\,Myr (Blaauw \cite{blaauw64}; Warren \& Hesser
\cite{warren78}; Brown, de Geus, \& de Zeeuw \cite{brown94}). This is
the age interval estimated for the O9.5V-type star based on its
physical properties, evolutionary stage (still burning hydrogen on the
main sequence) and membership in the Orion OB1b subgroup (Blaauw
\cite{blaauw91}). Other properties of the $\sigma$\,Orionis cluster,
e.g., distance (352\,pc) and reddening ($A_V\,\le\,0.5$\,mag), are
discussed in B\'ejar et al$.$ (\cite{bejar01a}). Here we examine low
mass stars and brown dwarfs with confirmed membership to determine the
most likely age of the cluster. We report on observations of
intermediate- and low-resolution optical spectroscopy in
Sects.~\ref{data} and~\ref{analysis}. A discussion and main
conclusions are given in Sects.~\ref{discussion}
and~\ref{conclusions}, respectively.


\begin{figure}
\centering
\includegraphics[width=8.8cm]{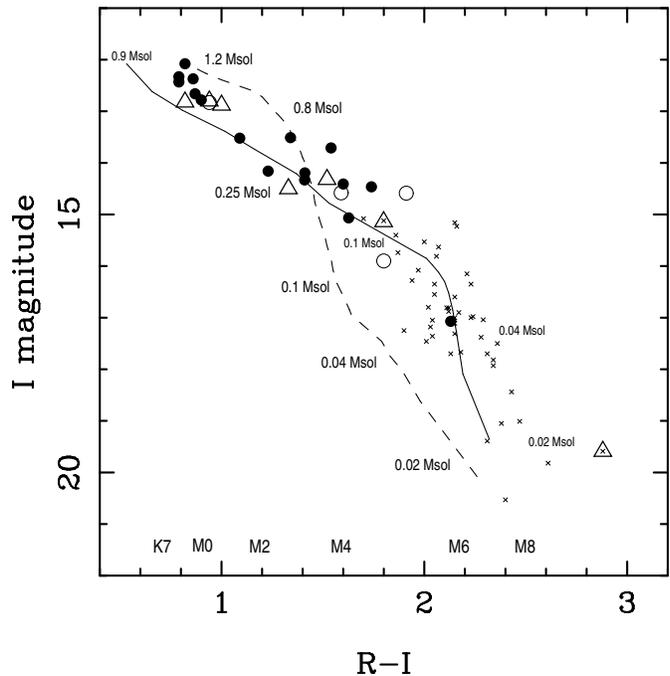}
\caption{\label{ri} Optical color-magnitude diagram of our 
  $\sigma$\,Orionis targets (large symbols). Open circles indicate
  sources with H$\alpha$ emissions larger than pEW\,=\,10\,\AA. Open
  triangles stand for cluster members with forbidden emission lines.
  S\,Ori objects (small crosses) from B\'ejar et al. (\cite{bejar99})
  are included in the figure for a better delineation of the
  $\sigma$\,Orionis cluster photometric sequence. Overplotted onto the
  data are the no-dust 5\,Myr isochrone (dashed line) of Baraffe et
  al$.$ (\cite{baraffe98}) and the 3\,Myr isochrone (solid line) of
  D'Antona \& Mazzitelli (\cite{dantona97}), for which masses in solar
  units have been labelled. Spectral types as a function of the
  $(R-I)$ color are also provided.}
\end{figure}

\begin{table*}
\caption[]{\label{log} Log of the observations (ordered by observing date).}
\begin{tabular}{lccccccl}
\hline
\noalign{\smallskip}
Object       & $\alpha$ ~(J2000)~ $\delta$                & $I$   & Obs. date   & Observ.      & Disp.   & Wl. range & Exp. time \\
             & ($^{h ~ m ~ s}$) ~~ ($^{\circ} ~ ' ~ ''$)  & (mag) &   (UT)      &            &(\AA/pix)& (\AA)     &   (s)     \\
\noalign{\smallskip}
\hline
\noalign{\smallskip}
4771--1075                & 05 39 05.3 ~ --2 32 30 & 12.66 & 20 Nov 1998 & CAHA & 0.55 & 6220--7297 & 700  \\
4771--1097                & 05 38 35.7 ~ --2 30 43 & 12.43 & 20 Nov 1998 & CAHA & 0.55 & 6220--7297 & 800  \\    
r053907--0228             & 05 39 07.6 ~ --2 28 23 & 14.33 & 20 Nov 1998 & CAHA & 0.55 & 6220--7297 & 2$\times$1200 \\ 
S\,Ori\,J053958.1--022619 & 05 39 58.1 ~ --2 26 19 & 14.19 & 21 Nov 1998 & CAHA & 0.55 & 6220--7297 & 2$\times$1200 \\ 
S\,Ori\,J053920.5--022737 & 05 39 20.5 ~ --2 27 37 & 13.51 & 21 Nov 1998 & CAHA & 0.55 & 6220--7297 & 2$\times$1200 \\ 
r053833-0236$^\ast$       & 05 38 33.9 ~ --2 36 38 & 13.71 & 21 Nov 1998 & CAHA & 0.55 & 6220--7297 & 2$\times$1500 \\
                          &                        &       & 22 Nov 1998 & CAHA & 2.41 & 6194--10000 & 650 \\
                          &                        &       & 26 Jan 1999 & ORM  & 0.84 & 6034--6840 & 1200 \\
S\,Ori\,J053949.3--022346 & 05 39 49.3 ~ --2 23 46 & 15.14 & 21 Nov 1998 & CAHA & 0.55 & 6220--7297 & 1800 \\ 
4771--1051$^\ast$         & 05 38 44.1 ~ --2 40 20 & 12.33 & 21 Nov 1998 & CAHA & 0.55 & 6220--7297 & 900, 535 \\
                          &                        &       & 26 Jan 1999 & ORM  & 0.84 & 6034--6840 & 1200 \\
S\,Ori\,J053715.1--024202 & 05 37 15.1 ~ --2 42 02 & 15.07 & 21 Nov 1998 & CAHA & 2.41 & 6194--10000& 600 \\ 
S\,Ori\,J053951.6--022248 & 05 39 51.6 ~ --2 22 48 & 14.59 & 22 Nov 1998 & CAHA & 2.41 & 6194--10000& 2$\times$600 \\ 
S\,Ori\,45                & 05 38 25.5 ~ --2 48 36 & 19.59 & 21 Dec 1998 &Keck& 0.85 & 6324--8025 & 3$\times$1800 \\
S\,Ori\,27                & 05 38 17.3 ~ --2 40 24 & 17.07 & 21 Dec 1998 &Keck& 0.85 & 6324--8025 & 2$\times$1200 \\
r053820--0237$^\ast$      & 05 38 20.3 ~ --2 37 47 & 12.83 & 25 Jan 1999 & ORM  & 0.84 & 6034--6993 & 2$\times$1200 \\
r053831--0235$^\ast$      & 05 38 31.4 ~ --2 35 15 & 13.52 & 25 Jan 1999 & ORM  & 0.84 & 6034--6840 & 600, 1800 \\
4771--899$^\ast$          & 05 38 47.9 ~ --2 27 14 & 12.08 & 25 Jan 1999 & ORM  & 0.84 & 6034--6840 & 300 \\
S\,Ori\,J053847.5--022711 & 05 38 47.5 ~ --2 27 11 & 14.46 & 26 Jan 1999 & ORM  & 0.84 & 6034--6840 & 2$\times$1800 \\ 
S\,Ori\,J054005.1--023052 & 05 40 05.1 ~ --2 30 52 & 15.90 & 26 Jan 1999 & ORM  & 0.84 & 6034--6840 & 2$\times$2700 \\ 
S\,Ori\,J054001.8--022133 & 05 40 01.8 ~ --2 21 33 & 14.32 & 26 Jan 1999 & ORM  & 0.84 & 6034--6840 & 2$\times$1200 \\ 
r053838-0236$^\ast$       & 05 38 38.0 ~ --2 36 38 & 12.37 & 27 Jan 1999 & ORM  & 0.84 & 6034--6840 & 2$\times$100 \\
4771--41                  & 05 38 27.1 ~ --2 45 10 & 12.82 & 27 Jan 1999 & ORM  & 0.84 & 6034--6840 & 2$\times$1200 \\
4771--1038$^\ast$         & 05 39 11.5 ~ --2 36 03 & 12.78 & 28 Jan 1999 & ORM  & 0.84 & 6034--6840 & 2$\times$2400 \\
r053840-0230$^\ast$       & 05 38 40.2 ~ --2 30 19 & 12.80 & 28 Jan 1999 & ORM  & 0.84 & 6034--6840 & 2$\times$2400 \\
r053820--0234             & 05 38 20.4 ~ --2 34 09 & 14.58 & 29 Jan 1999 & ORM  & 0.84 & 6034--6840 & 900, 1800 \\ 
r053849--0238$^\ast$      & 05 38 49.0 ~ --2 38 21 & 12.88 & 03 Dec 1999 &McDonald &0.70&6150--6850 & 5$\times$1200 \\
r053923--0233$^\ast$      & 05 39 22.7 ~ --2 33 33 & 14.16 & 03 Dec 1999 &McDonald &0.70&6150--6850 & 8$\times$1200 \\
S\,Ori\,J053827.4--023504 & 05 38 27.4 ~ --2 35 04 & 14.50 & 05 Dec 1999 &McDonald &0.70&6150--6850 &12$\times$1200 \\ 
S\,Ori\,J053914.5--022834 & 05 39 14.5 ~ --2 28 34 & 14.75 & 06 Dec 1999 &McDonald &0.70&6150--6850 & 2$\times$1200 \\ 
S\,Ori\,J053820.1--023802 & 05 38 20.1 ~ --2 38 02 & 14.41 & 06 Dec 1999 &McDonald &0.70&6150--6850 &10$\times$1200 \\ 
\hline
\noalign{\smallskip}
\end{tabular}
\\
$^\ast$ ~ Also detected in X-rays (Wolk \cite{wolk96}).
\end{table*}

\begin{table*}
\caption[]{\label{setup} Instrumental setups of different campaigns.}
\begin{tabular}{cccccccc}
\hline
\noalign{\smallskip}
\multicolumn{1}{c}{Run} &
\multicolumn{1}{c}{Observ.} &
\multicolumn{1}{c}{Teles.} &
\multicolumn{1}{c}{Spectrograph} &
\multicolumn{1}{c}{Grating} &
\multicolumn{1}{c}{Detector} &
\multicolumn{1}{c}{Spatial res.} &
\multicolumn{1}{c}{Slit width} \\
\noalign{\smallskip}
\hline
\noalign{\smallskip}
20 Nov 1998 & CAHA   & 3.5-m & TWIN & T06      & SITe 2000$\times$800\,pix & 0.56\arcsec/pix & 1.2\arcsec \\
21 Nov 1998 & CAHA   & 3.5-m & TWIN & T11      & SITe 2000$\times$800\,pix & 0.56\arcsec/pix & 1.2\arcsec \\
20 Dec 1998 & Keck   & Keck II& LRIS & 900/5500 & 2048$\times$2048\,pix     & 0.22\arcsec/pix & 1.0\arcsec \\
25--28 Jan 1999 & ORM & INT  & IDS  & R1200Y  & Tektronix 1024$\times$1024\,pix & 0.70\arcsec/pix & 1.7\arcsec \\
03--06 Dec 1999 & McDonald & 2.1-m  & ES2 & \#25 & Craf/Cassini 1024$\times$1024\,pix & 2.72\arcsec/pix & 2.1\arcsec \\
\noalign{\smallskip}
\hline
\end{tabular}
\end{table*}

\section{Sample selection}
Our list of 28 targets (12.3\,$\le$\,$I$\,$\le$19.6, $T_{\rm
  eff}$\,$\sim$\,4200--2400\,K) comprises $\sigma$\,Orionis solar-mass
and low mass stars, and brown dwarfs selected from different surveys
(see Table~\ref{log}). All have been identified as genuine cluster
members using various techniques. Stars labelled with ``4771'' and
``r'' were first identified by Wolk (\cite{wolk96}). $VRIJHK$
photometry, spectroscopy, and strong X-ray detections (in many cases)
are available. Wolk (\cite{wolk96}) provided equivalent widths of the
Li\,{\sc i} resonance doublet for a few of these stars. However, his
spectroscopic data of relatively faint sources have poor
signal-to-noise (S/N) ratios, which severely affects the measurements.
We decided to re-observe them to improve the quality of the spectra.
S\,Ori targets (IAU nomenclature) have been selected from the $RIJ$
survey of B\'ejar (\cite{bejar01b}). They nicely fit in the cluster
optical-infrared sequence. Albeit we lack previous spectroscopic data
for them, the spectra presented here confirm them as very active,
young sources, and therefore, they have to be cluster members. The two
brown dwarfs in our sample, S\,Ori\,27 and 45, have been taken from
B\'ejar et al$.$ (\cite{bejar99}), where they are discussed at length.

\begin{figure}
\centering
\includegraphics[width=8.8cm]{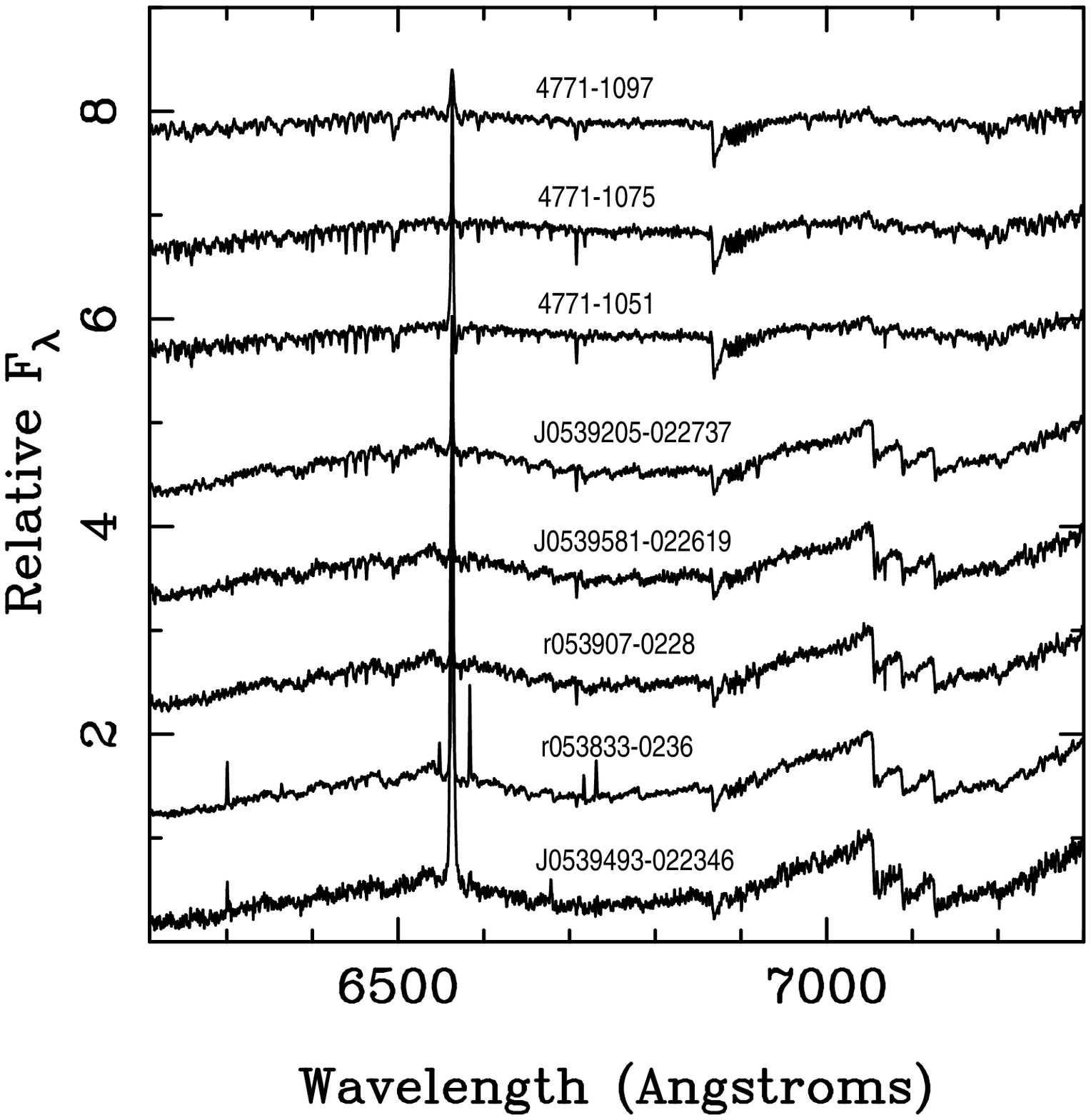}
\includegraphics[width=8.8cm]{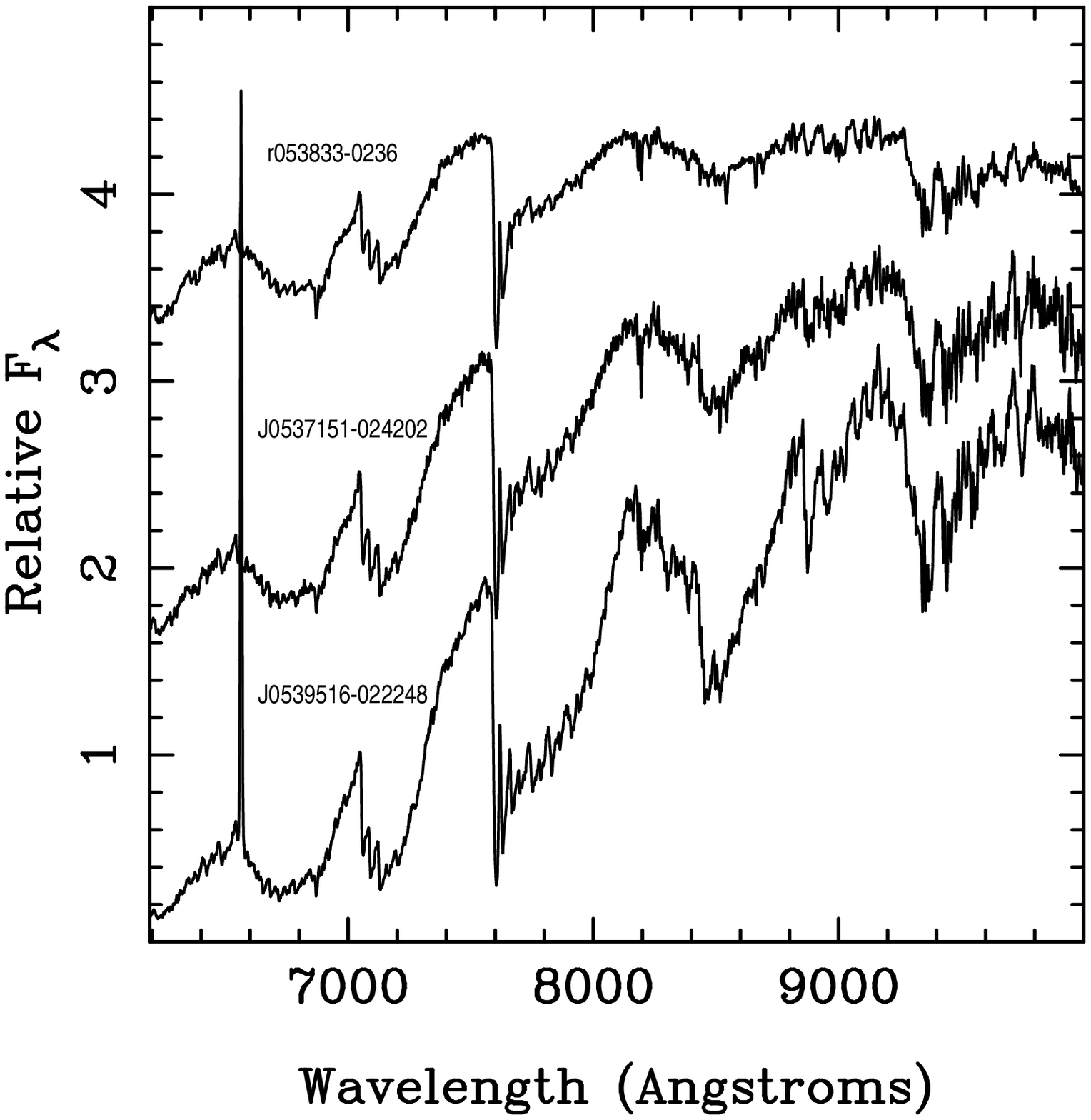}
\caption{\label{calar} CAHA spectra. 
  Data have been normalized to the counts at around 7050\,\AA~and
  shifted by 1 in the upper panel and by 1.5 in the lower panel for
  clarity.}
\end{figure}

\begin{figure}
  \centering \includegraphics[width=8.8cm]{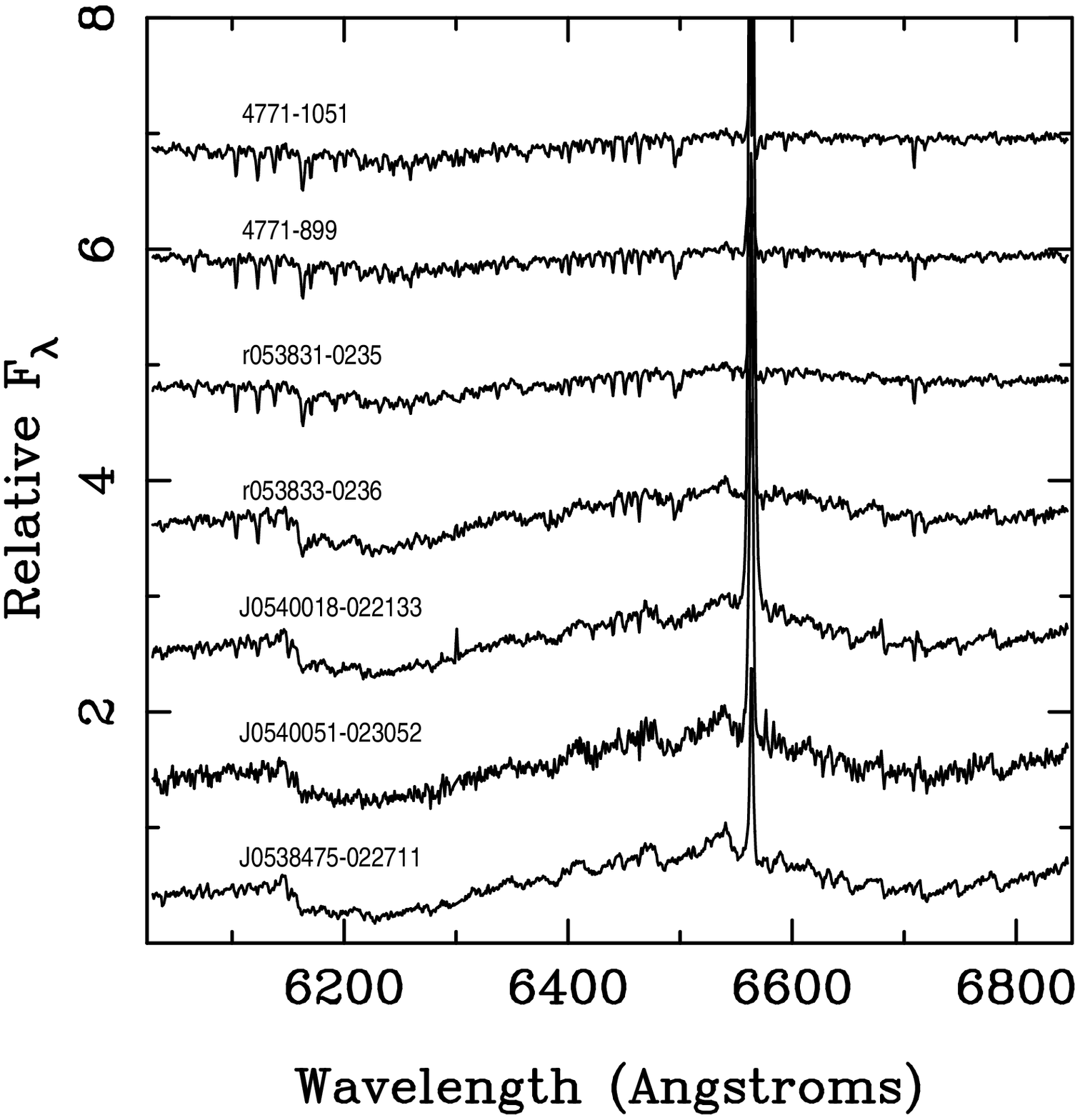}
  \includegraphics[width=8.8cm]{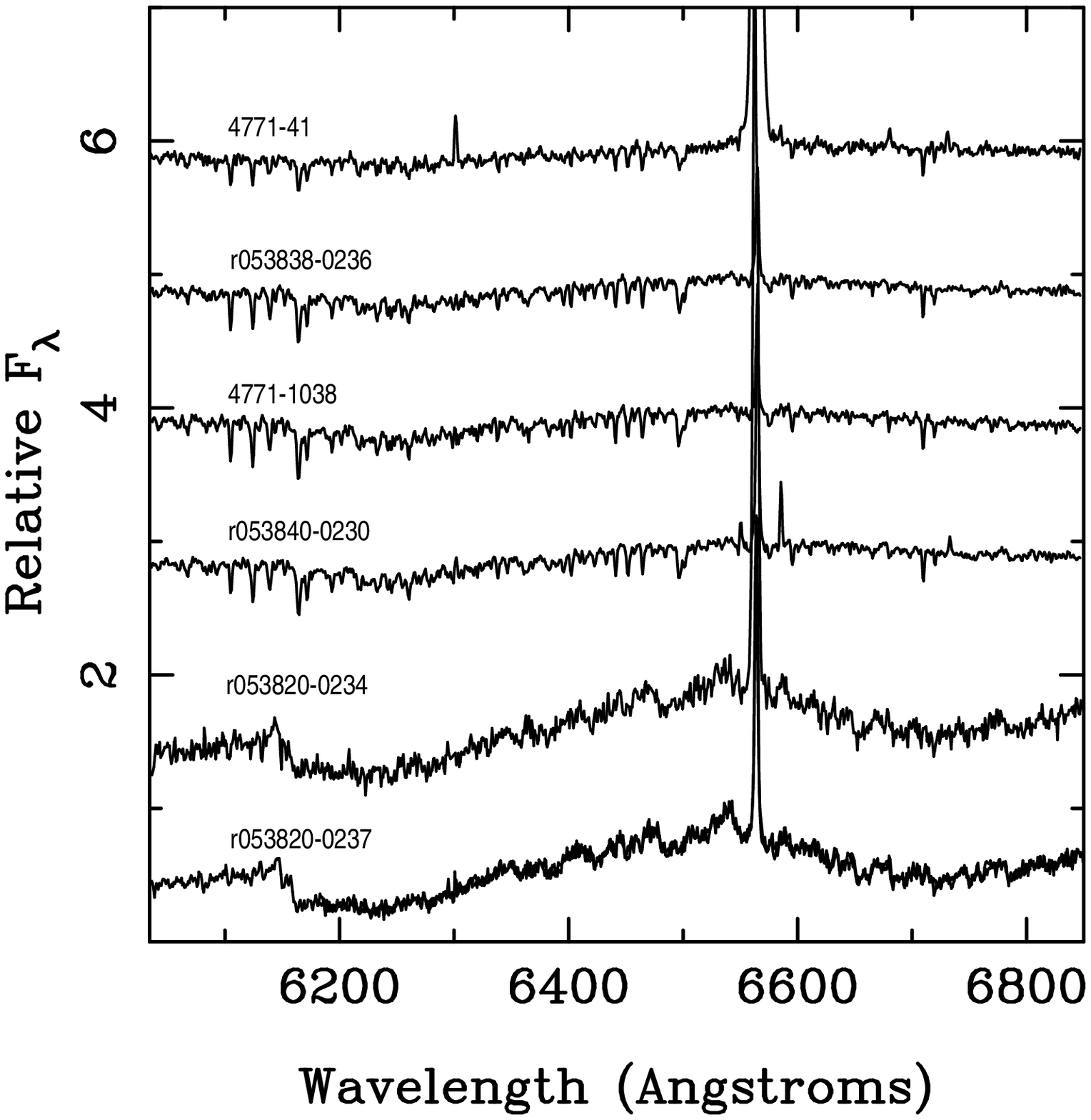}
\caption{\label{int} ORM spectra. 
  Data have been normalized to the counts at around 6535\,\AA~and
  shifted by 1 for clarity.}
\end{figure}

Names, coordinates and $I$ magnitudes of our sample are provided in
Table~\ref{log}. Their location in the optical color-magnitude diagram
is illustrated in Fig.~\ref{ri}, where $RI$ photometry has been taken
from Wolk (\cite{wolk96}), B\'ejar et al. (\cite{bejar99}) and B\'ejar
(\cite{bejar01b}). Overplotted are the solar metallicity, and
``no-dust'' 3\,Myr and 5\,Myr isochrones of D'Antona \& Mazzitelli
(\cite{dantona97}) and Baraffe et al$.$ (\cite{baraffe98}),
respectively. Masses as predicted by the models are indicated in the
figure. Comparisons with other tracks are provided in B\'ejar et al$.$
(\cite{bejar99}). Our targets have masses ranging from
1.2\,$M_{\odot}$ down to roughly 0.02\,$M_{\odot}$.


\begin{figure}
\centering
\includegraphics[width=8.8cm]{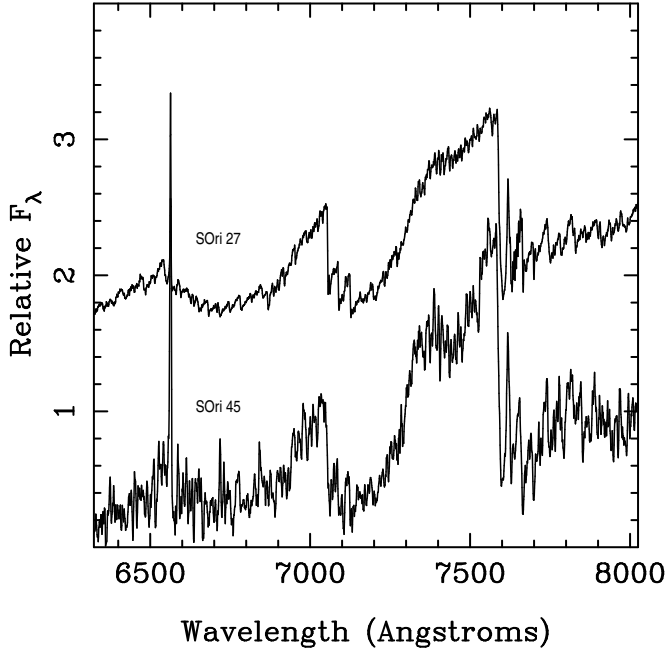}
\caption{\label{keck} Keck spectra of two brown 
  dwarfs in the $\sigma$\,Orionis cluster. Data have been normalized
  to the counts at around 7050\,\AA~and shifted by 1.5 for clarity.
  The spectrum of S\,Ori\,45 has been smoothed with a boxcar of 5
  pixels to increase the S/N ratio of the data.}
\end{figure}

\begin{figure}
\centering
\includegraphics[width=8.8cm]{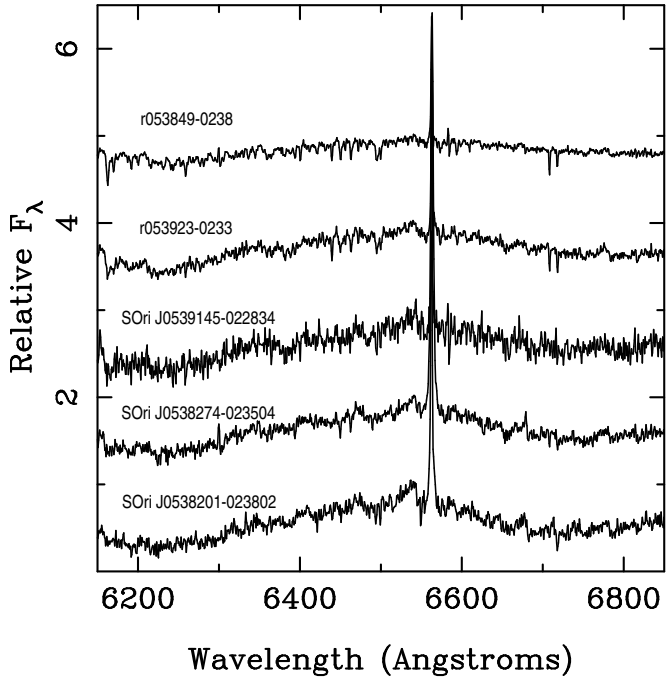}
\caption{\label{mcdonald} McDonald spectra. 
  Data have been normalized to the counts at around 6540\,\AA~and
  shifted by 1 for clarity.}
\end{figure}

\section{Observations and data reduction \label{data}}
We acquired intermediate- to low-resolution optical spectra using the
following telescopes: the 3.5-m telescope at Calar Alto (CAHA)
Observatory in Almer\'\i a (Spain), the 2.5-m Isaac Newton telescope
(INT) at the Roque de los Muchachos Observatory (ORM) on the island of
La Palma (Spain), the 10-m Keck II telescope at Mauna Kea Observatory
on the island of Hawaii (U.S.A.), and the Otto Struve 2.1-m telescope
at McDonald Observatory in west Texas (U.S.A.). Observing campaigns,
spectrographs attached to the Cassegrain focus of each telescope,
detectors, gratings and slit widths used for collecting data are
summarized in Table~\ref{setup}. The red arm of the TWIN instrument
(CAHA) and the 235\,mm camera of the IDS spectrograph (ORM) were
chosen. Table~\ref{log} shows the journal of the observations, which
includes the nominal dispersions of the instrumental setups. No
binning of the pixels along the spectral direction and the projection
of the slits onto the detectors yielded spectral resolutions of
1.54\,\AA~($R$\,$\sim$\,4400, CAHA, first run),
4.82\,\AA~($R$\,$\sim$\,1600, CAHA, second run),
1.68\,\AA~($R$\,$\sim$\,3800, ORM), 2.89\,\AA~($R$\,$\sim$\,2500,
Keck), 1.4\,\AA~($R$\,$\sim$\,4600, McDonald), and spatial resolutions
as listed in Table~\ref{setup}. A binning of 8 pixels along the
spatial direction was applied to the CCD at McDonald. Filters blocking
the light blueward of 5000\AA~were used at the CAHA and ORM
telescopes. No order-blocking filter was used at the Keck II
telescope; nevertheless, the two targets observed are very red and the
contribution of their blue light to the far-red optical spectrum is
negligible (Mart\'\i n et al. \cite{martin99}). Weather conditions
during the four runs (CAHA, ORM, Keck and McDonald) were
spectroscopic. The seeing in the visible was stable at around
1\arcsec~at Keck, and 1.5\arcsec--2\arcsec~at CAHA and ORM. The
spatial resolution at McDonald was 2.72\arcsec/pix due to some
technical problems related to the instrumentation.

Raw images were reduced with standard procedures including bias
subtraction and flat-fielding within {\sc noao iraf}\footnote{IRAF is
  distributed by National Optical Astronomy Observatory, which is
  operated by the Association of Universities for Research in
  Astronomy, Inc., under contract with the National Science
  Foundation.}. We extracted object and sky spectra using the optimal
extraction algorithm available in the {\sc apextract} package. A full
wavelength solution from calibration lamps taken immediately after
each target was applied to the spectra. The $rms$ of the fourth-order
polynomial fit to the wavelength calibration is typically 5--10\%~the
nominal dispersion. To complete the data reduction, we corrected the
extracted spectra for instrumental response using data of
spectrophotometric standard stars (HD\,19445, Feige\,34, G\,191\,B2B,
BD+26\,2606) obtained on the same nights and with the same
instrumental configurations. These stars have fluxes available in the
{\sc iraf} environment (Massey et al. \cite{massey88}).

\begin{figure}
\centering
\includegraphics[width=8.8cm]{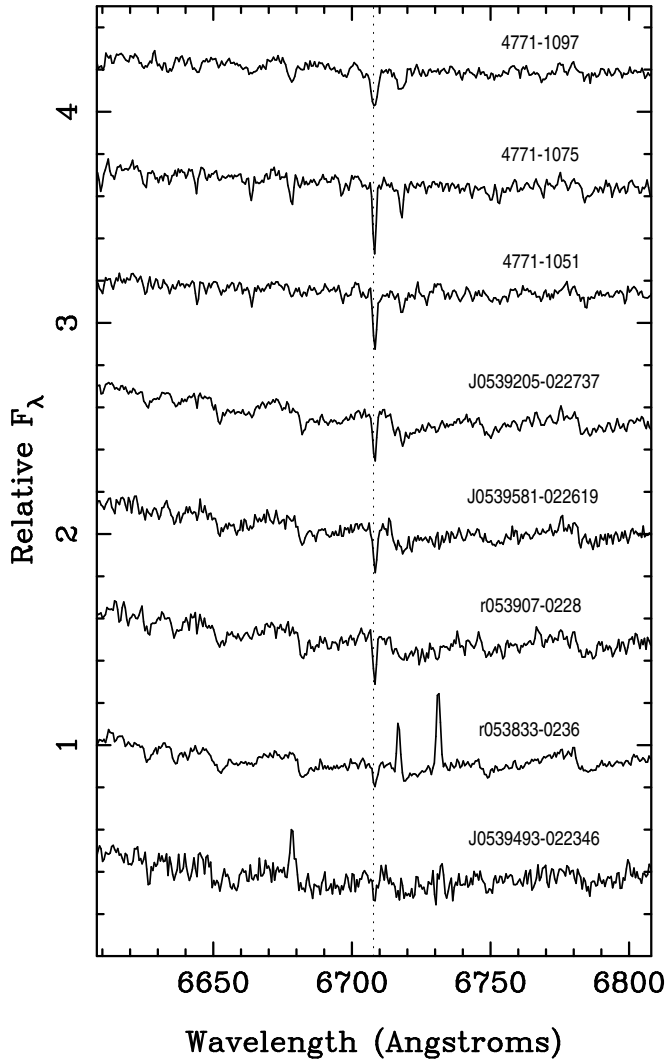}
\caption{\label{licalar} Region around the Li\,{\sc i} 
  $\lambda$6708\,\AA~line (CAHA high-resolution spectra). The star
  4771--1097 is a fast rotator.}
\end{figure}

\begin{figure}
\centering
\includegraphics[width=8.8cm]{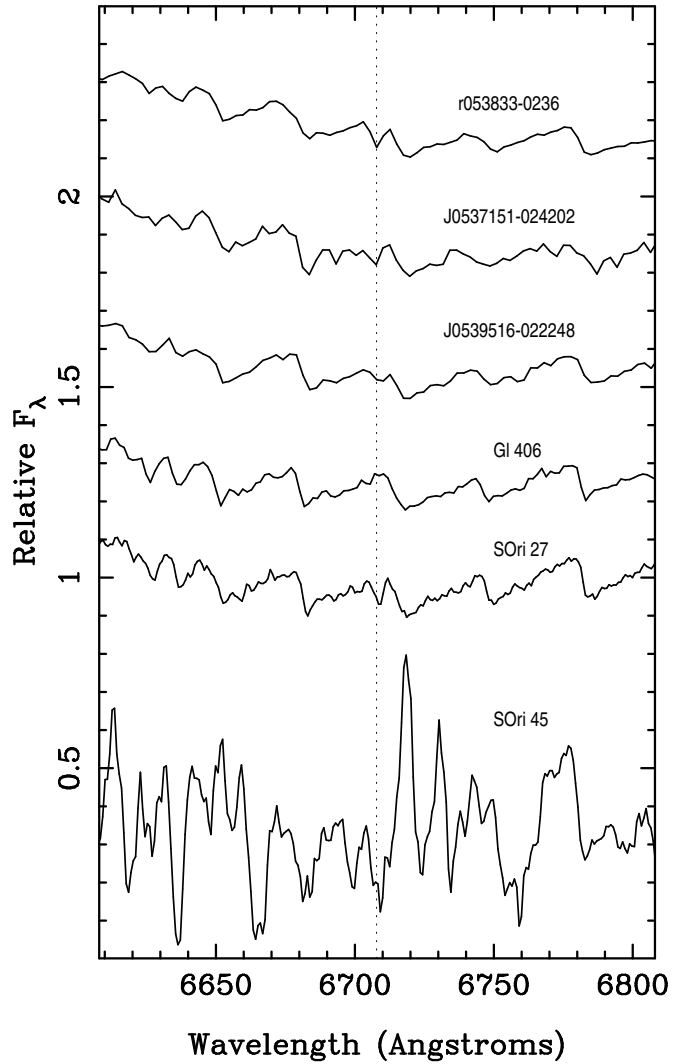}
\caption{\label{licakeck} Region around the 
  Li\,{\sc i} $\lambda$6708\,\AA~line (CAHA and Keck spectra). The
  spectrum of the M6-type field star Gl\,406 is included for
  comparison. Data have been shifted by different constants for
  clarity.}
\end{figure}

\begin{figure}
\centering
\includegraphics[width=8.8cm]{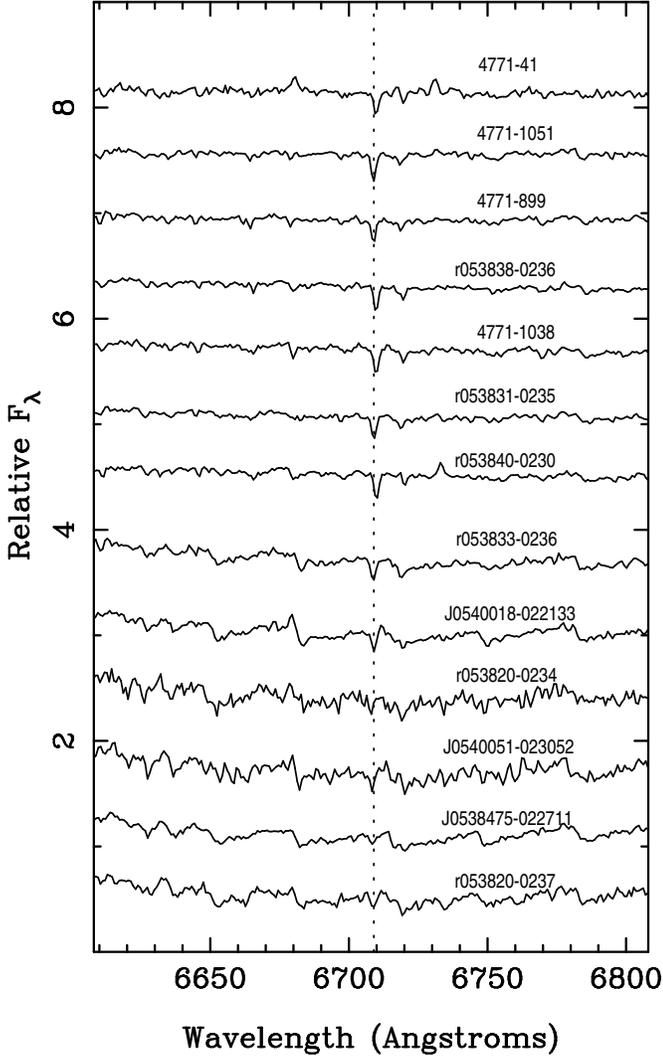}
\caption{\label{liint} Region around the 
  Li\,{\sc i} $\lambda$6708\,\AA~line (ORM spectra). Data have been
  shifted by 0.6 for clarity.}
\end{figure}

\begin{figure}
\centering
\includegraphics[width=8.8cm]{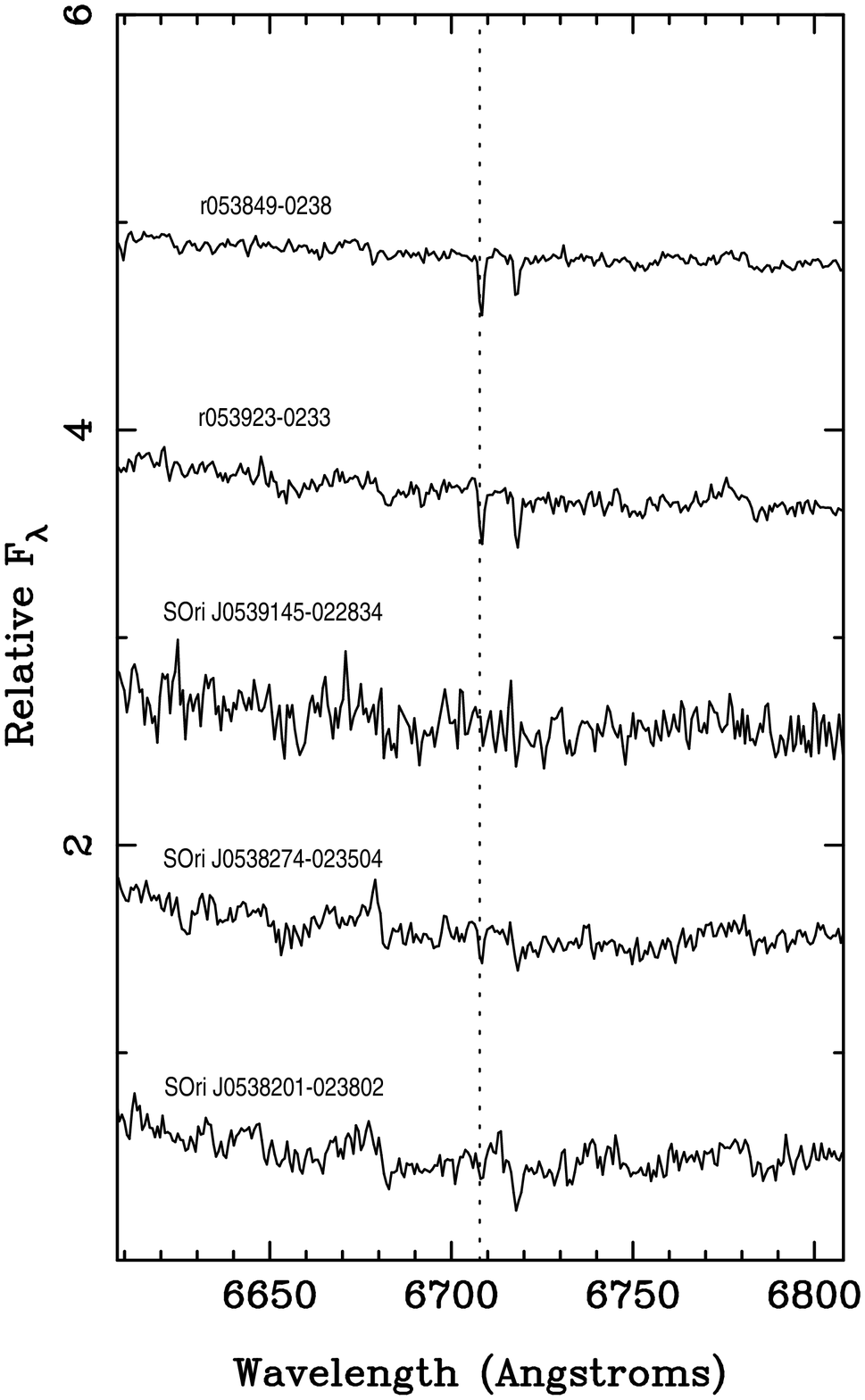}
\caption{\label{limcdonald} Region around 
  the Li\,{\sc i} $\lambda$6708\,\AA~line (McDonald spectra).}
\end{figure}

The resulting spectra are depicted in
Figs.~\ref{calar}--\ref{mcdonald}. They are ordered by increasingly
late spectral type and shifted by a constant for clarity. The region
around the Li\,{\sc i} $\lambda$6708\,\AA~line is amplified in
Figs.~\ref{licalar}--\ref{limcdonald}. In Fig.~\ref{licakeck} we have
included the spectrum of the field M6-type spectroscopic standard star
Gl\,406 for a better comparison.

\section{Analysis and results \label{analysis}}
\subsection{Spectral types}
We inferred spectral types by comparing our target spectra to data of
spectroscopic standard stars (Gl\,820A, K5V; Gl\,820B, K7V; Gl\,338A,
M0V; Gl\,182, M0.5V; Gl\,767A, M1V; Gl\,767B, M2.5V; Gl\,569A, M3V;
Gl\,873, M3.5; Gl\,402, M4V; Gl\,905, M5V; and Gl\,406, M6V). The
reference spectra were obtained with similar instrumentations in
previous campaigns. In addition, we observed several K- and M-type
standards with the McDonald telescope. For the spectral
classification, we also used molecular indices that are based on the
relative strengths of CaH and TiO bands (Kirkpatrick, Henry \&
McCarthy \cite{kirk91}; Prosser, Stauffer \& Kraft \cite{prosser91}),
and the pseudocontinuum PC3 index given in Mart\'\i n, Rebolo \&
Zapatero Osorio (\cite{martin96}), which is valid for types later than
M3. Our measurements, with an uncertainty of half a subclass, are
provided in Table~\ref{ew}. The spectral types of S\,Ori\,27 and 45
have been taken from B\'ejar et al$.$ (\cite{bejar99}). The final
adopted spectral classes are in the range K6--M8.5.

\begin{table*}
\caption[]{\label{ew} Spectroscopic data.}
\begin{tabular}{lccccrcccc}
\hline
\noalign{\smallskip}
\multicolumn{1}{c}{Object$^a$} &
\multicolumn{1}{c}{$I$}    & 
\multicolumn{1}{c}{$R-I$}  & 
\multicolumn{1}{c}{Sp.\,T.$^b$}& 
\multicolumn{1}{c}{MJD$^c$}& 
\multicolumn{1}{c}{pEW$^d$ (H$\alpha$)} & 
\multicolumn{1}{c}{pEW$^e$ (Li\,{\sc i})}   &
\multicolumn{1}{c}{log\,$L_{\rm H\alpha}/L_{\rm bol}$} &
\multicolumn{1}{c}{$v_{r}$} &
\multicolumn{1}{c}{Template} \\
\multicolumn{1}{c}{      } &
\multicolumn{1}{c}{   }    & 
\multicolumn{1}{c}{     }  & 
\multicolumn{1}{c}{       }& 
\multicolumn{1}{c}{(--51000)}& 
\multicolumn{1}{c}{(\AA)             } & 
\multicolumn{1}{c}{(\AA)           }   &
\multicolumn{1}{c}{       }& 
\multicolumn{1}{c}{(km\,s$^{-1}$)} &
\multicolumn{1}{c}{       } \\
\noalign{\smallskip}
\hline
\noalign{\smallskip}
4771--1075        & 12.66 & 0.87 & K7.0 & 137.9536 &  0.7$\pm$0.1 & 0.59$\pm$0.09 & --4.25$\pm$0.09 & 29.7$\pm$7 & 4771--1051 \\
4771--1097        & 12.43 & 0.79 & K6.0 & 137.9686 &  2.2$\pm$0.8 & 0.47$\pm$0.07 & --3.77$\pm$0.19 & 34.9$\pm$7 & 4771--1051 \\
r053907--0228     & 14.33 & 1.41 & M3.0 & 137.9881 &  3.6$\pm$0.7 & 0.67$\pm$0.08 & --3.63$\pm$0.13 & 39.0$\pm$7 & 4771--1051 \\
J053958.1--022619 & 14.19 & 1.41 & M3.0 & 138.0676 &  4.0$\pm$0.8 & 0.73$\pm$0.10 & --3.59$\pm$0.13 & 47.4$\pm$7 & 4771--1051 \\
J053920.5--022737 & 13.51 & 1.34 & M2.0 & 138.1046 &  3.2$\pm$0.7 & 0.65$\pm$0.10 & --3.66$\pm$0.14 & 33.0$\pm$7 & 4771--1051 \\
r053833--0236     & 13.71 & 1.54 & M4.0 & 138.1237 & 14.0$\pm$2.0 & 0.60$\pm$0.10 & --3.10$\pm$0.11 & 35.8$\pm$7 & 4771--1051 \\
                  &       &      & M3.0 & 139.1788 &  2.7$\pm$0.3 & 0.47$\pm$0.05 & --3.82$\pm$0.10 & 35.8$\pm$7 & -- \\
                  &       &      & M3.5 & 204.0726 &  2.2$\pm$0.8 & 0.61$\pm$0.08 & --3.90$\pm$0.19 & 35.6$\pm$10& 4771--1051 \\
J053949.3--022346 & 15.14 & 1.80 & M4.0 & 138.1696 & 42.0$\pm$6.0 & 0.71$\pm$0.15 & --2.78$\pm$0.11 & 38.0$\pm$7 & 4771--1051 \\
4771-1051         & 12.33 & 0.79 & K7.5 & 138.2073 &  6.4$\pm$1.0 & 0.58$\pm$0.09 & --3.31$\pm$0.12 & 32.8$\pm$7 & Gl\,14     \\
                  &       &      & K8.0 & 204.0912 &  5.5$\pm$1.0 & 0.59$\pm$0.05 & --3.37$\pm$0.13 & 32.8$\pm$7 & -- \\
J053715.1--024202 & 15.07 & 1.63 & M4.0 & 138.9739 &  4.9$\pm$0.5 & 0.42$\pm$0.08 & --3.61$\pm$0.10 & 35.9$\pm$25& r053833--0236 \\
J053951.6--022248 & 14.59 & 1.91 & M5.5 & 139.0881 & 60.0$\pm$7.0 & 0.35$\pm$0.10 & --2.71$\pm$0.10 & 40.5$\pm$25& r053833--0236 \\
S\,Ori\,45        & 19.59 & 2.88 & M8.5 & 168.4644 & 33.0$\pm$9.0 & 2.40$\pm$1.00 & --3.74$\pm$0.16 &--13.0$\pm$15 & vB\,10   \\
S\,Ori\,27        & 17.07 & 2.13 & M6.5 & 168.5276 &  5.7$\pm$0.5 & 0.74$\pm$0.09 & --3.89$\pm$0.09 & 35.5$\pm$10& vB\,10     \\
r053820--0237     & 12.83 & 0.94 & M5.0 & 203.8482 & 10.2$\pm$0.8 & 0.66$\pm$0.10 & --3.09$\pm$0.09 & 50.7$\pm$10& 4771--1051 \\
r053831--0235     & 13.52 & 1.09 & M0.0 & 203.8849 &  4.5$\pm$0.5 & 0.47$\pm$0.07 & --3.44$\pm$0.10 & 35.1$\pm$10& 4771--1051 \\
4771--899         & 12.08 & 0.82 & K7.0 & 203.9329 &  3.1$\pm$0.5 & 0.48$\pm$0.07 & --3.61$\pm$0.12 & 31.0$\pm$10& 4771--1051 \\
J053847.5--022711 & 14.46 & 1.74 & M5.0 & 203.9476 &  7.8$\pm$1.0 & 0.53$\pm$0.08 & --3.47$\pm$0.11 & 40.5$\pm$10& 4771--1051 \\
J054005.1--023052 & 15.90 & 1.80 & M5.0 & 204.0101 & 20.5$\pm$6.0 & 0.72$\pm$0.15 & --3.09$\pm$0.17 & 33.6$\pm$10& 4771--1051 \\
J054001.8--022133 & 14.32 & 1.52 & M4.0 & 204.0382 & 46.5$\pm$9.0 & 0.65$\pm$0.15 & --2.57$\pm$0.13 & 41.9$\pm$10& 4771--1051 \\
r053838--0236     & 12.37 & 0.86 & K8.0 & 205.9079 &  2.9$\pm$0.5 & 0.53$\pm$0.05 & --3.68$\pm$0.12 & 41.6$\pm$10& 4771--1051 \\
4771--41          & 12.82 & 0.82 & K7.0 & 205.9222 & 53.5$\pm$9.0 & 0.50$\pm$0.06 & --2.38$\pm$0.12 & 47.7$\pm$10& 4771--1051 \\
4771--1038        & 12.78 & 0.90 & K8.0 & 206.0002 &  2.0$\pm$0.5 & 0.58$\pm$0.09 & --3.79$\pm$0.15 & 38.7$\pm$10& 4771--1051 \\
r053840--0230     & 12.80 & 0.94 & M0.0 & 206.0299 &  6.7$\pm$0.6 & 0.52$\pm$0.05 & --3.27$\pm$0.09 & 46.3$\pm$10& 4771--1051 \\
r053820--0234     & 14.58 & 1.59 & M4.0 & 207.0720 & 28.0$\pm$4.0 & 0.45$\pm$0.15 & --2.83$\pm$0.11 & 47.7$\pm$10& 4771--1051 \\
r053849--0238     & 12.88 & 1.00 & M0.5 & 515.2615 &  2.6$\pm$0.3 & 0.55$\pm$0.05 & --3.67$\pm$0.10 & 29.0$\pm$10& Gl\,873, Gl\,182 \\
r053923--0233     & 14.16 & 1.23 & M2.0 & 515.3933 &  4.1$\pm$0.5 & 0.54$\pm$0.08 & --3.51$\pm$0.10 & 31.0$\pm$10& Gl\,873, Gl\,182 \\
J053827.4--023504 & 14.50 & 1.33 & M3.5 & 517.4297 & 21.2$\pm$3.0 & 0.52$\pm$0.05 & --2.83$\pm$0.11 & 36.7$\pm$10& Gl\,873, Gl\,182 \\
J053914.5--022834 & 14.75 & 1.48 & M3.5 & 518.2596 &  4.2$\pm$0.7 & $\le$0.44     & ---             & 31.3$\pm$10& Gl\,873, Gl\,182 \\
J053820.1--023802 & 14.41 & 1.60 & M4.0 & 518.3899 &  9.6$\pm$2.0 & 0.57$\pm$0.07 & --3.30$\pm$0.14 & 29.2$\pm$10& Gl\,873, Gl\,182 \\
\hline
\noalign{\smallskip}
\end{tabular}
\\
$^a$ ~ Note the drop of ``S\,Ori'' for some objects. \\
$^b$ ~ Uncertainty of half a subclass.\\
$^c$ ~ Modified Julian date at the beginning of the exposure.\\
$^d$ ~ In emission. Whenever more than one spectrum available, 
 the pEW has been measured over the combined data.\\
$^e$ ~ In absorption. Whenever more than one spectrum available, 
 the pEW has been measured over the combined data.
\end{table*}

We note that our spectral classification relies on field dwarf objects
with high gravities. The gravity of $\sigma$\,Orionis cluster members
is expected to be around log\,$g$\,=\,4.0 (CGS units) according to the
evolutionary models of Baraffe et al$.$ (\cite{baraffe98}) and
D'Antona \& Mazzitelli (\cite{dantona94}). Older K-type stellar
counterparts in the field ($\sim$5\,Gyr) display similar gravities,
but early-M and late-M stars have values 0.5\,dex and 1.0\,dex larger,
respectively. Cool giants are characterized by very low gravities
(log\,$g$\,=\,1.5--2, Bonnell \& Tell \cite{bonnell93}; van Belle
\cite{vanbelle99}). Therefore, it is reasonable to base the spectral
classification of young late-type objects on a scheme intermediate
between that of dwarfs and that of giants. Luhman (\cite{luhman99})
successfully applied this exercise to members of the young cluster
IC\,348, inferring that the spectral classification of objects like
those of $\sigma$\,Orionis can be obtained from dwarfs with an
accuracy up to half a subclass. We have confirmed this by comparing
the optical spectrum of our M8.5 brown dwarf with brown dwarfs of
identical types in $\rho$\,Oph and IC\,348 (Luhman, Liebert \& Rieke
\cite{luhman97}; Luhman \cite{luhman99}). The three spectra overlap
very nicely. We are confident that the spectral types given in
Table~\ref{ew} are reliable within the quoted uncertainty.

\begin{figure}
\centering
\includegraphics[width=8.8cm]{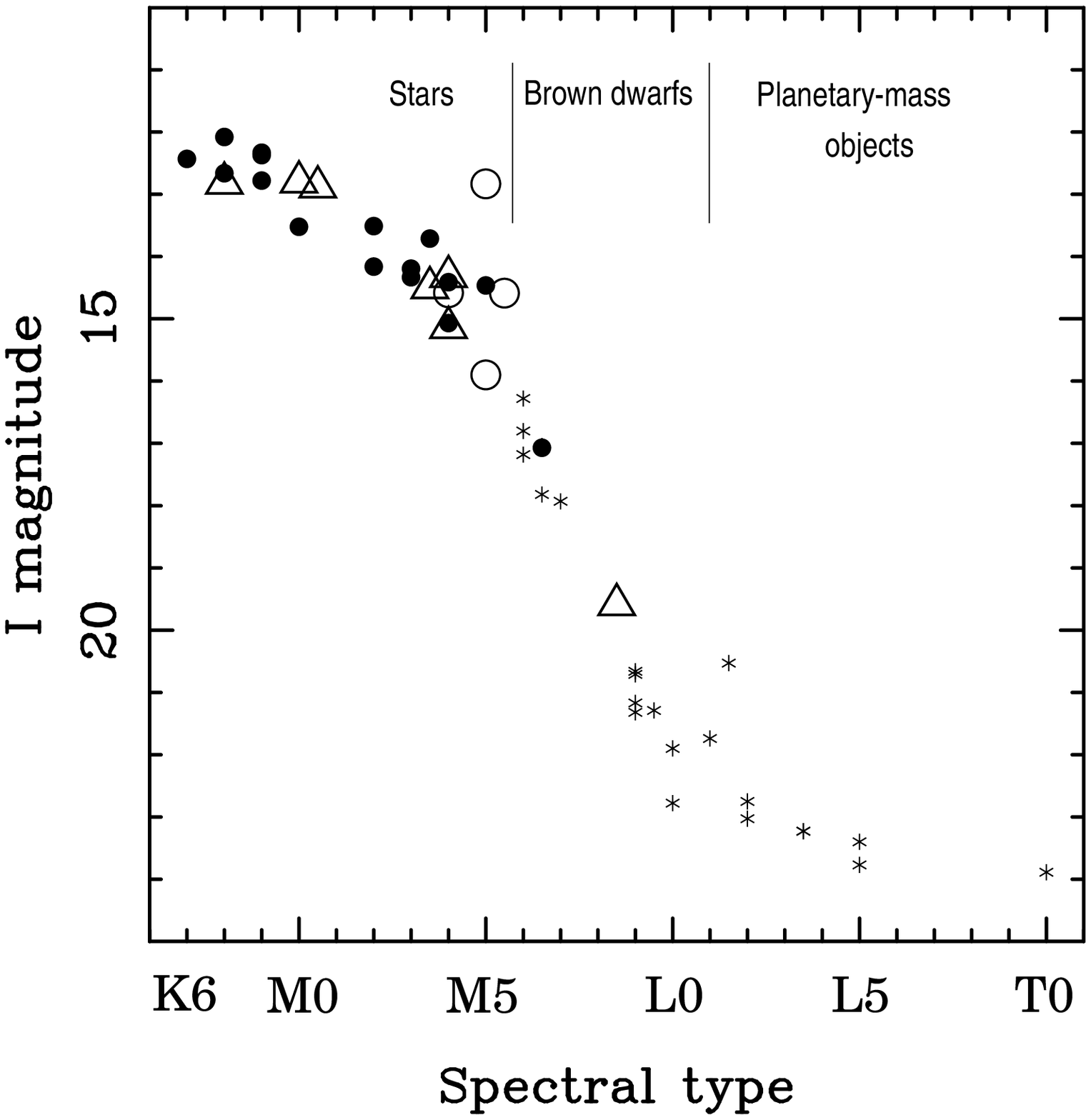}
\caption{\label{tspmag} $I$ magnitude against spectral 
  type for $\sigma$\,Orionis members. Symbols are as in Fig.~\ref{ri}.
  The lower ``end'' of the sequence (asterisks) is completed with data
  taken from B\'ejar et al. (\cite{bejar99}), Barrado y Navascu\'es et
  al. (\cite{barrado01a}) and Mart\'\i n et al$.$ (\cite{martin01}).
  Typical uncertainties in spectral type are half a subclass, except
  for the coolest objects ($\ge$L2), where an uncertainty of one
  subclass is expected.}
\end{figure}

Since the spectral classification reflects effective temperatures,
cluster members should lie along a defined sequence in magnitude {\em
  vs} spectral type diagrams. The $\sigma$\,Orionis spectroscopic
sequence is depicted in Fig.~\ref{tspmag}, where we have combined data
presented here with data taken from B\'ejar et al$.$ (\cite{bejar99}),
Barrado y Navascu\'es et al$.$ (\cite{barrado01a}) and Mart\'\i n et
al$.$ (\cite{martin01}). We note that the figure covers a wide range
of masses: stars, brown dwarfs and planetary-mass objects.
Free-floating low mass stars and isolated planetary-mass objects in
the $\sigma$\,Orionis cluster have luminosities in the $I$-band that
differ by about 3 orders of magnitude. Because substellar objects
contract and fade very rapidly, such a difference becomes incredibly
large at older ages, e.g., 8 orders of magnitude at 100\,Myr (Chabrier
et al$.$ \cite{chabrier00}).

\subsection{Rotational velocities}
Given the poor velocity resolution of our spectra (68\,km\,s$^{-1}$,
CAHA, first run; 176\,km\,s$^{-1}$, CAHA, second run;
78\,km\,s$^{-1}$, ORM; 120\,km\,s$^{-1}$, Keck; and 65\,km\,s$^{-1}$,
McDonald), we are able to detect extremely fast rotators. These are
defined as objects with projected rotational velocities, $v\,{\rm
  sin}\,i$, larger than 55\,km\,s$^{-1}$. The only case in
Table~\ref{ew} is 4771--1097, which was observed with the largest
dispersion. This star (K6) is the most massive object in our sample
($\sim$0.9--1.2\,$M_{\odot}$). We measured the rotational velocity by
comparing its spectrum to a slowly rotating template selected from our
sample. The spectrum of 4771--1075 (observed with the same
instrumentation) has very sharp lines and a similar spectral type, as
can be seen from Fig.~\ref{licalar}. The full width at half-maximum of
the atomic lines of this ``reference'' star indicates that its
spectral broadening is mainly instrumental. Our procedure was to
produce a set of artificial spectra spinned up to velocities of 75,
80, 95 and 110\,km\,s$^{-1}$.  We then compared the observed spectrum
of 4771--1097 to the synthetic rotational spectra and performed an
analysis using the minimum squares technique. Spectral regions free of
emission lines and telluric absorptions were considered. We derived
$v\,{\rm sin}\,i$\,=\,80$\pm$15\,\,km\,s$^{-1}$. Wolk (\cite{wolk96})
found clear evidence of optical photometric variability in 4771--1097
and inferred a likely rotational period of $\sim$1\,day.  Combining
these results with predicted radii for masses in the range
0.9--1.2\,$M_{\odot}$ and ages between 3 and 5\,Myr (D'Antona \&
Mazzitelli \cite{dantona94}; Baraffe et al$.$ \cite{baraffe98}), we
conclude that 4771--1097 is rotating with an inclination of
$i$\,=\,50$^{\circ}$--90$^{\circ}$.

\subsection{Radial velocities}
We computed radial velocities via Fourier cross-correlation of the
target spectra with templates of similar spectral type. Our CAHA and
ORM measurements were calibrated with the radial velocity standard
star Gl\,14 ($v_{r}$\,=\,3.3$\pm$0.3\,km\,s$^{-1}$, Marcy \& Benitz
\cite{marcy89}; Marcy \& Chen \cite{marcy92}), which was observed with
the largest dispersion at CAHA on 1998 Nov. 20. We used this star to
derive the radial velocity of 4771--1051, and then correlated the rest
of our CAHA targets of the same resolution against it. We also adopted
4771--1051 as the template for the ORM data. Whether this star has a
variable radial velocity is unknown to us. Thus, our ORM radial
velocities might be shifted by a certain amount. However, this is
unlikely (at least within the error bars of the measurements) since
there is another $\sigma$\,Orionis member, r053833--0236, observed at
CAHA and ORM. After correlating the ORM spectra of these two stars, we
obtained for r053833--0236 a heliocentric radial velocity similar to
the one derived from the CAHA data. We note that the relative radial
velocity of each target with respect 4771--1051 is reliable. We
adopted r053833--0236 as the reference star for the low-resolution
CAHA spectra, and the M8 field star vB\,10 was used as the template
($v_{r}$\,=\,35.3$\pm$1.5\,km\,s$^{-1}$, Tinney \& Reid
\cite{tinney98}) for the Keck spectra. The spectrum of vB\,10 was
taken from Mart\'\i n et al$.$ (\cite{martin96}). Our McDonald spectra
were cross-correlated against the radial velocity standard stars
Gl\,182 ($v_{r}$\,=\,32.4$\pm$1.5\,km\,s$^{-1}$, Jeffries
\cite{jeffries95}) and Gl\,873
($v_{r}$\,=\,0.47$\pm$0.24\,km\,s$^{-1}$, Marcy, Lindsay \& Wilson
\cite{marcy87}), which were observed with the same instrumentation and
on the same nights.

Radial velocities, their uncertainties and the templates used are
provided in Table~\ref{ew}. We took special care in cross-correlating
spectral windows (e.g$.$ 6100--6800\,\AA, 8400--8800\,\AA) that are
not affected by telluric absorptions and that contain many
photospheric lines. In addition, we considered only parts of the
spectra free of emission lines. The error bars in the table point to a
possible 1/4 pixel uncertainty in the Fourier cross-correlation
technique (Mart\'\i n et al$.$ \cite{martin99}; Lane et al$.$
\cite{lane01}). We have checked this by cross-correlating the McDonald
spectra against two reference stars. The spectrum of S\,Ori\,45 is
rather noisy, and the quoted error bar comes from the dispersion
observed after cross-correlating different spectral regions. The
majority of our radial velocities are obtained to an accuracy of the
order of 10\,km\,s$^{-1}$. After discarding the largest and smallest
radial velocity values from Table~\ref{ew} (i.e., r053820--0237 and
S\,Ori\,45, respectively), the mean heliocentric radial velocity of
our $\sigma$\,Orionis sample is $<v_r>$\,=\,37.3\,km\,s$^{-1}$ with a
standard deviation of 5.8\,km\,s$^{-1}$. This is comparable to the
systemic radial velocity of the cluster's central star, which has been
determined to be in the range 27--38\,km\,s$^{-1}$ (Bohannan \&
Garmany \cite{bohannan78}; Garmany et al$.$ \cite{garmany80}; Morrell
\& Levato \cite{morrell91}). Additionally, these velocities (except
for one, see Sect$.$~\ref{discussion}) are consistent with our sample
belonging to the Orion OB association (Alcal\'a et al$.$
\cite{alcala00}), and their distribution is significantly different
from that of field stars.

\subsection{H$\alpha$ emission}
We derived H$\alpha$ pseudo-equivalent widths via direct integration
of the line profile with the task {\sc splot} in {\sc iraf}. We note
that given the cool nature of our sample, equivalent widths in the
optical are generally measured relative to the observed local
pseudo-continuum formed by (mainly TiO) molecular absorptions
(Pavlenko \cite{pav97}). We will refer to these equivalent widths as
``pseudo-equivalent widths'' (pEWs).

Because of the resolution of our observations, broad H$\alpha$ lines
appear blended with other nearby spectral features. The results of our
measurements, given in Table~\ref{ew}, have been extracted by adopting
the base of the line as the continuum. The error bars were obtained
after integrating over the reasonable range of possible continua.
Although this procedure does not give an absolute equivalent width,
i.e., measured with respect the real continuum, it is commonly used by
various authors, and allows us to compare our values with those
published in the literature. We note that all of our program objects
show H$\alpha$ in emission and that no significant H$\alpha$
variability is found in any of them, except for r053833--0236 and
S\,Ori\,45. We also note that the H$\alpha$ emission of the fast
rotator 4771--1097 is not stronger than that of other similar-type
cluster members.

\begin{figure}
\centering
\includegraphics[width=8.8cm]{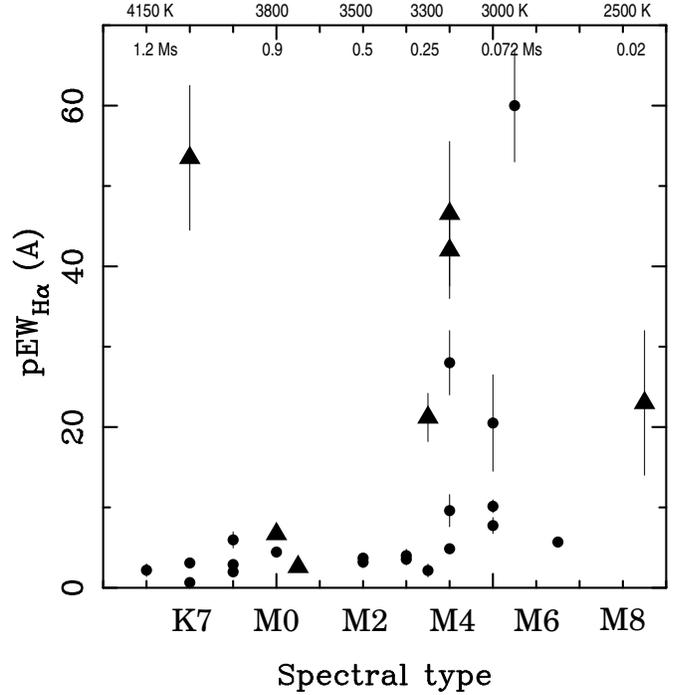}
\caption{\label{ha} Pseudo-equivalent widths of H$\alpha$ emission 
  as a function of spectral type (Table~\ref{ew}). Objects with other
  emission lines are plotted with filled triangles, except for
  r053833--0236 (M3.5).  Typical uncertainty in spectral type is half
  a subclass. The stellar-substellar borderline takes place at M5--M6
  at the age of the cluster. Effective temperatures in Kelvin and
  masses in solar units are also given.}
\end{figure}

Figure~\ref{ha} shows the distribution of H$\alpha$ pEWs as a function
of spectral type. Effective temperatures are given on the basis of the
temperature---spectral-type relationships by Leggett et al$.$
(\cite{leggett96}), Jones et al$.$ (\cite{jones95}) and Bessell
(\cite{bessell91}). Masses as inferred from the 5\,Myr evolutionary
isochrone of Baraffe et al$.$ (\cite{baraffe98}) are also indicated in
the figure. In general, there is a trend of increasing H$\alpha$
emission for cooler spectral classes, i.e., for lower masses. This
behavior has been observed in various young clusters, like the
Pleiades and Hyades (Stauffer et al$.$ \cite{stauffer94}), IC\,4665
(Prosser \cite{prosser93}), $\alpha$\,Persei (Prosser
\cite{prosser94}), and Praesepe (Barrado y Navascu\'es, Stauffer \&
Randich \cite{barrado98}). The relative increase of H${\alpha}$ in
M-dwarfs may be (at least partially) explained by the drop of the flux
continuum and the larger TiO molecular absorptions in the optical as a
consequence of cooler $T_{\rm eff}$s. We note that, on average,
H$\alpha$ for a given spectral type is slightly larger in
$\sigma$\,Orionis than in other open clusters. This is very likely a
direct consequence of the marked youth of $\sigma$\,Orionis.

\begin{figure}
\centering
\includegraphics[width=8.8cm]{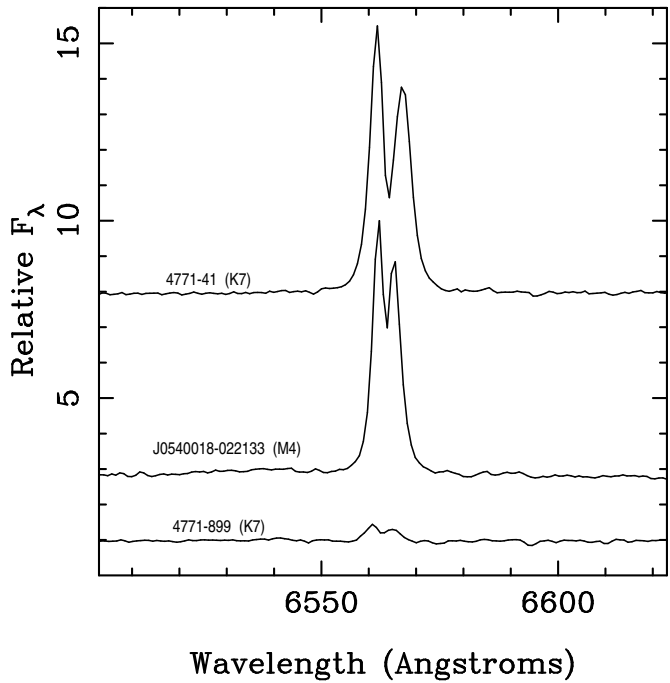}
\caption{\label{hadp} Double-peak H$\alpha$ profiles. From top 
  to bottom, the base of the emission line spreads over $\sim
  \pm$400\,km\,s$^{-1}$, $\pm$350\,km\,s$^{-1}$ and
  $\pm$270\,km\,s$^{-1}$.}
\end{figure}

In Fig.~\ref{ha} H$\alpha$ emission appears very strong
(pEWs\,$\ge$\,20\,\AA) and dispersed for late spectral classes
($\ge$M3.5), corresponding to masses below 0.25\,$M_{\odot}$ at the
age of $\sigma$\,Orionis. Various authors have found an apparent
``turnover'' in the distribution of H$\alpha$ emission in the Pleiades
(Stauffer et al$.$ \cite{stauffer94}; Hodgkin, Jameson \& Steele
\cite{hodgkin95}) and $\alpha$\,Persei (Zapatero Osorio et al$.$
\cite{osorio96}). Pleiades and $\alpha$\,Per stars with spectral types
later than M3.5--M4 show a lower level of emission than stars with
warmer classes. The authors suggest that this turnover is due to the
transition from radiative to convective cores. By inspecting D'Antona
\& Mazzitelli (\cite{dantona94}) pre-main sequence evolutionary
models, we find that this transition takes place at masses
0.3--0.2\,$M_{\odot}$ regardless of age. In $\sigma$\,Orionis we do
not see a drop in the H$\alpha$ emission of fully convective objects,
but an enhacement. The source of such large emission clearly
diminishes by the age of the $\alpha$\,Persei cluster (90\,Myr,
Stauffer et al$.$ \cite{stauffer99}). However, the emission level of
more massive stars remains with similar strengths.

\begin{figure}
\centering
\includegraphics[width=8.8cm]{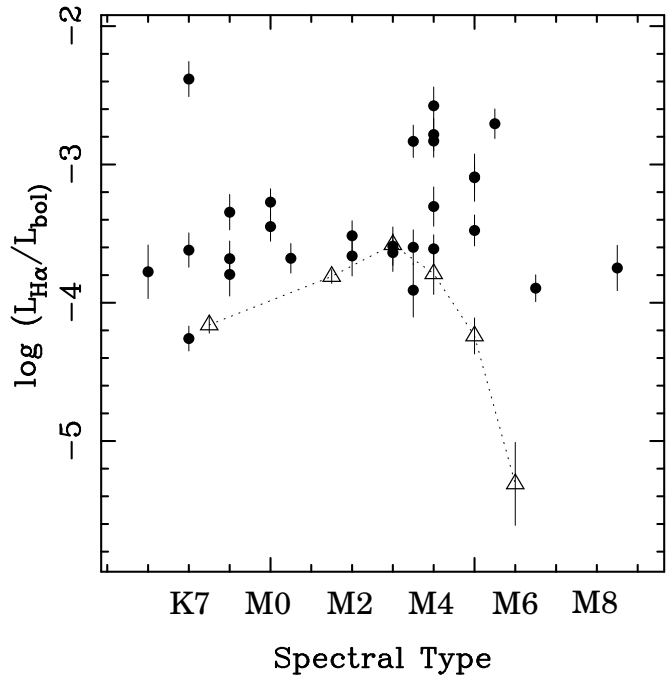}
\caption{\label{habol} Ratio of H$\alpha$ luminosity of the 
  object to its bolometric luminosity as a function of optical
  spectral type. $\sigma$\,Orionis members are plotted with filled
  circles. For comparison we have also indicated Pleiades mean values
  with open triangles and a dotted line (Hodgkin et al$.$
  \cite{hodgkin95}). Uncertainty in spectral type is half a subclass.}
\end{figure}

Three stars in our sample, namely 4771--41 (K7),
S\,Ori\,J054001.8--022133 (M4) and 4771--899 (K7), show profiles of
H$\alpha$ emission similar to those of classical T\,Tauri (CTT) stars,
i.e., double peak structure and very broad lines spanning over
$\pm$300\,km\,s$^{-1}$ from the line center. We illustrate in
Fig.~\ref{hadp} the region around H$\alpha$ for these objects. While
the emission intensity is rather large in 4771--41 and
S\,Ori\,J054001.8--022133 (pEWs above 45\,\AA), it is moderate in
4771--899.

\begin{table*}
\caption[]{\label{prohibidas} Pseudo-equivalent widths (pEWs) of emission lines.}
\begin{tabular}{lccccccccc}
\hline
\noalign{\smallskip}
\multicolumn{2}{c}{ } &
\multicolumn{2}{c}{[O\,{\sc i}]} &
\multicolumn{2}{c}{[N\,{\sc ii}]} &
\multicolumn{1}{c}{ } &
\multicolumn{1}{c}{He\,{\sc i} } &
\multicolumn{2}{c}{[S\,{\sc ii}]} \\
\multicolumn{1}{c}{Object} &
\multicolumn{1}{c}{MJD$^a$}    &
\multicolumn{1}{c}{$\lambda$6300\,\AA} &
\multicolumn{1}{c}{$\lambda$6364\,\AA} &
\multicolumn{1}{c}{$\lambda$6548\,\AA} &
\multicolumn{1}{c}{$\lambda$6583\,\AA} &
\multicolumn{1}{c}{H$\alpha$} &
\multicolumn{1}{c}{$\lambda$6678\,\AA} &
\multicolumn{1}{c}{$\lambda$6716\,\AA} &
\multicolumn{1}{c}{$\lambda$6731\,\AA} \\
\multicolumn{1}{c}{ } &
\multicolumn{1}{c}{(--51000) } &
\multicolumn{1}{c}{(\AA)} &
\multicolumn{1}{c}{(\AA)} &
\multicolumn{1}{c}{(\AA)} &
\multicolumn{1}{c}{(\AA)} &
\multicolumn{1}{c}{(\AA)} &
\multicolumn{1}{c}{(\AA)} &
\multicolumn{1}{c}{(\AA)} &
\multicolumn{1}{c}{(\AA)} \\
\noalign{\smallskip}
\hline
\noalign{\smallskip}
r053833--0236             & 138.1237 & 2.29$\pm$0.10 & 0.75$\pm$0.05 & 0.80$\pm$0.05 & 2.50$\pm$0.10 & 14.0$\pm$2.0 & $\le$0.1      & 1.15$\pm$0.08 & 1.65$\pm$0.08 \\
                          & 138.1433 & 2.00$\pm$0.10 & 0.57$\pm$0.05 & 0.88$\pm$0.05 & 2.54$\pm$0.10 & 14.1$\pm$2.0 & 0.15$\pm$0.05 & 0.85$\pm$0.05 & 1.49$\pm$0.05 \\
J053949.3--022346         & 138.1696 & 1.80$\pm$0.50 & $\le$0.2      & $\le$1.0      & 0.35$\pm$0.08 & 42.0$\pm$6.0 & 1.18$\pm$0.08 & $\le$0.15     & $\le$0.15     \\
S\,Ori\,45$^b$            & 168.4644 & --            & $\le$1.5      & $\le$1.5      &  5.0$\pm$3.0  & 33.0$\pm$9.0 & $\le$1.5      & 5.0$\pm$3.0   & 3.56$\pm$2.0 \\
J054001.8--022133         & 204.0382 & 1.27$\pm$0.10 & 0.33$\pm$0.05 & 0.26$\pm$0.05 & 0.33$\pm$0.05 & 47.0$\pm$9.0 & 1.05$\pm$0.08 & $\le$0.15     & $\le$0.15     \\
                          & 204.0531 & 0.45$\pm$0.10 & $\le$0.1      & 0.10$\pm$0.05 & 0.15$\pm$0.05 & 46.0$\pm$9.0 & 0.38$\pm$0.08 & $\le$0.15     & $\le$0.15     \\
4771--41                  & 205.9222 & 0.96$\pm$0.10 & $\le$0.1      & blended       & 0.25$\pm$0.05 & 53.2$\pm$9.0 & 0.21$\pm$0.05 & 0.16$\pm$0.08 & 0.42$\pm$0.05 \\
                          & 205.9366 & 1.00$\pm$0.10 & 0.18$\pm$0.08 & $\le$0.1      & 0.30$\pm$0.05 & 54.0$\pm$9.0 & 0.42$\pm$0.07 & $\le$0.1      & 0.45$\pm$0.05 \\
r053840--0230             & 206.0299 & $\le$0.15     & $\le$0.15     & 0.42$\pm$0.05 & 1.10$\pm$0.10 &  6.5$\pm$0.6 & $\le$0.1      & $\le$0.15     & 0.31$\pm$0.05 \\
                          & 206.0582 & 0.55$\pm$0.05 & 0.15$\pm$0.05 & 0.54$\pm$0.05 & 1.10$\pm$0.10 &  6.9$\pm$0.6 & $\le$0.1      & $\le$0.15     & 0.31$\pm$0.05 \\
r053849--0238             & 515.2615 & 0.36$\pm$0.10 & $\le$0.1      & $\le$0.1      & 0.32$\pm$0.05 &  2.6$\pm$0.3 & $\le$0.1      & $\le$0.1      & $\le$0.1   \\
J053827.4--023504 & 517.4297 & 1.25$\pm$0.50 & $\le$0.3      & $\le$0.3      & $\le$0.3      & 21.2$\pm$3.0 & 0.26$\pm$0.08 & $\le$0.3      & $\le$0.3   \\
\hline
\noalign{\smallskip}
\end{tabular}
\\
$^a$ ~ Modified Julian date at the beginning of the exposure.\\
$^b$ ~ Measures over the combined spectrum. Individual H$\alpha$ 
 pEWs were 20.0, 49.4 and 22.0$\pm$7.0\,\AA, respectively.
\end{table*}

We have calculated the H$\alpha$ luminosity ($L_{\rm H\alpha}$) for
our sample as in Herbst \& Miller (\cite{herbst89}) and Hodgkin et
al$.$ (\cite{hodgkin95}). The ratio of $L_{\rm H\alpha}$ to bolometric
luminosity ($L_{\rm H\alpha}/L_{\rm bol}$) is independent of the
surface area and represents the fraction of the total energy output in
H$\alpha$. To derive $L_{\rm bol}$ we have used bolometric corrections
provided by Monet et al$.$ (\cite{monet92}) and Kenyon \& Hartmann
(\cite{kenyon95}). The logarithmic values of $L_{\rm H\alpha}/L_{\rm
  bol}$ are listed in Table~\ref{data}; uncertainties take into
account errors in photometry and in H$\alpha$ pEWs. Figure~\ref{habol}
shows the distribution of log\,($L_{\rm H\alpha}/L_{\rm bol}$) with
spectral type. For comparison purposes, we have also included the
Pleiades mean values (Hodgkin et al$.$ \cite{hodgkin95}). In the
Pleiades, the $L_{\rm H\alpha}/L_{\rm bol}$ ratio clearly increases to
a maximum at around the M3 spectral type and then turns over. This is
not observed in the $\sigma$\,Orionis cluster, where cooler objects
present larger H$\alpha$ output fluxes than the older Pleiades
spectral counterparts. Discarding $\sigma$\,Orionis members with
log\,($L_{\rm H\alpha}/L_{\rm bol}$)\,$\ge$\,--3.2\,dex, cluster data
appear to display a flat distribution from late K to late M (i.e., no
dependence on color and mass) at around log\,($L_{\rm H\alpha}/L_{\rm
  bol}$)\,=\,--3.61\,dex, with a standard deviation of 0.18\,dex.

\subsection{Other emission lines}
Our targets are pre-main sequence objects characterized by significant
H$\alpha$ emission and the presence of lithium in their atmospheres
(see Sect$.$~\ref{lithium}). All show properties that resemble
T\,Tauri stars. The nominal definition of weak-lined T\,Tauri (WTT)
stars is usually based on H$\alpha$ emission: pEWs smaller than
10\,\AA~for K and early-M stars (Herbig \& Bell \cite{herbig88}) and
smaller than 20\,\AA~for later M-types (Mart\'\i n \cite{martin98}).
This is accomplished by many of our objects.

Some of our program targets display, however, other permitted
(He\,{\sc i} $\lambda$6678\,\AA) and forbidden ([O\,{\sc i}]
$\lambda$6300\,\AA, [N\,{\sc ii}] $\lambda$6548, $\lambda$6583\,\AA,
[S\,{\sc ii}] $\lambda$6716, $\lambda$6731\,\AA) emission lines. We
have measured their pEWs; values are given in Table~\ref{prohibidas}
as a function of Julian date. We note that some contamination from
terrestrial night-sky emission lines may be expected in the
measurements of the faintest sources. The objects of
Table~\ref{prohibidas} are plotted with different symbols in various
figures of this paper, except for r053833--0236 (for this star we have
used the ``quiet'' ORM data). The majority of the targets from Wolk
(\cite{wolk96}) are, in addition, classified as strong X-ray emitters
by this author. In contrast to the younger CTT stars, WTT objects are
not accreting mass from disks. However, the presence of He\,{\sc i}
and [O\,{\sc i}], [N\,{\sc ii}], [S\,{\sc ii}] emission lines is
related to jets and outflows, which are typical of CTT stars and
accretion processes (Edwards et al$.$ \cite{edwards87}; Hartigan,
Edwards \& Ghandour \cite{hartigan95}). These lines are generally
detected in objects with strong H$\alpha$ emissions
(pEWs\,$\ge$\,10\,\AA, see Fig.~\ref{ha}). The coexistence of
$\sigma$\,Orionis members with properties of WTT and CTT stars is
indeed indicative of ages of a few Myr. It may also indicate that
small objects are accreting for longer periods than are more massive
stars (Hillenbrand et al$.$ \cite{hillenbrand98}; Haisch, Lada \& Lada
\cite{haisch01}), provided that their strong H$\alpha$ emissions are
due to disk accretion.

The star r053833--0236 shows strong H$\alpha$ emission and noticeable
forbidden lines of [O\,{\sc i}], [N\,{\sc ii}] and [S\,{\sc ii}] in
two consecutive CAHA spectra (Fig.~\ref{calar}, upper panel). However,
its H$\alpha$ intensity clearly decreased, and no other emission lines
were present in data collected on the following night
(Fig.~\ref{calar}, lower panel) or with the INT (Fig.~\ref{int}). The
sources of this episodic flarelike event are not continuous in
r053833--0236, probably indicating inhomogeneus mass infall onto the
star surface.

The case of the brown dwarf S\,Ori\,45 ($\sim$0.02\,$M_{\odot}$) is
particularly interesting and noteworthy. Albeit the detection of
[N\,{\sc ii}] and [S\,{\sc ii}] emission lines is affected by large
uncertainties because of the modest quality of the Keck spectrum, this
finding is very encouraging. It suggests that substellar objects, even
those with very low masses, can sustain surrounding disks from which
matter is accreted. Muzerolle et al$.$ (\cite{muzerolle00}) has
recently reported on the evidence for disk accretion in a T\,Tauri
object at the substellar limit. The presence of disks around brown
dwarfs in the Trapezium cluster ($\sim$1\,Myr) has been proved by
Muench et al$.$ (\cite{muench01}). The emission lines observed in
S\,Ori\,45 indicate that ``substellar'' disks can last up to ages like
those of the $\sigma$\,Orionis cluster. It is also feasible that the
probable binary nature of S\,Ori\,45 (see Sect$.$~\ref{discussion})
triggers the formation of these emission lines. Nevertheless, further
spectroscopic data will be very valuable to confirm the presence of
forbidden emission lines in S\,Ori\,45. The rapid H$\alpha$
variability of this brown dwarf is also remarkable.

\subsection{Li\,{\sc i} absorption}
\subsubsection{Synthetic spectra}
We have computed theoretical optical spectra in the wavelength range
6680--6735\,\AA~around the Li\,{\sc i} $\lambda$6708\,\AA~resonance
doublet for gravity log\,$g$\,=\,4.0 (CGS units) and for $T_{\rm
  eff}$\,=\,4000--2600\,K by running the WITA6 code described in
Pavlenko (\cite{pav00}). This code is designed to opperate in the
framework of classical approximations: local thermodynamic equilibrium
(LTE), a plane-parallel geometry, neither sources nor drops of energy.
The synthetical spectra have been obtained using the atmospheric
  structure of the NextGen models published in Hauschildt, Allard, \&
Baron (\cite{haus99}).  We have adopted a microturbulent velocity
value of $v_t$\,=\,2\,km\,s$^{-1}$, solar elemental abundances (Anders
\& Grevesse \cite{anders89}), except for lithium, and solar isotopic
ratios for titanium and oxygen atoms. The ionization-dissociation
equilibria were solved for about 100 different species, where
constants of chemical equilibrium were taken from Tsuji
(\cite{tsuji73}) and Gurvitch et al$.$ (\cite{gurvitch79}). For the
particular case of the TiO molecule, we have adopted a dissociation
potential of $D_0$\,=\,7.9\,eV and the molecular line list of Plez
(\cite{plez98}). The atomic line parameters have been taken from the
VALD database (Piskunov et al$.$ \cite{piskunov95}), and the procedure
for computing damping constants is discussed in Pavlenko et al$.$
(\cite{pav95}) and Pavlenko (\cite{pav01}).

Synthetic spectra were originally obtained with a step of 0.03\,\AA~in
wavelength, and were later convolved with appropiate Gaussians to
match a resolution of 1.68\,\AA, which corresponds to the majority of
our data. We have produced a grid of theoretical spectra for nine
different abundances of lithium [log\,$N$(Li)\,=\,1.0, 1.3, ..., 3.1,
3.4, referred to the usual scale of log\,$N$(H)\,=\,12] and seven
values of $T_{\rm eff}$ (4000, 3600, 2400, 3200, 3000, 2800 and
2600\,K), covering the spectral sequence of our program targets.
Determinations of the meteoritic lithium abundance (Nichiporuk \&
Moore \cite{nichiporuk74}; Grevesse \& Sauval \cite{grevesse98}) lie
between log\,$N$(Li)\,=\,3.1 and 3.4. Extensive lithium studies
performed in solar metallicity, intermediate-age clusters like the
Pleiades (Soderblom et al$.$ \cite{soderblom93}), $\alpha$\,Per
(Balachandran, Lambert \& Stauffer \cite{balachandran96}), Blanco~1
(Jeffries \& James \cite{jeffries99}), NGC\,2516 (Jeffries, James \&
Thurston \cite{jeffries98}), and IC\,2602 and IC\,2391 (Randich et
al$.$ \cite{randich01}), as well as in the Taurus star-forming region
(Mart\'\i n et al$.$ \cite{martin94}) show that non-depleted stars
preserve an amount of lithium compatible with a logarithmic abundance
between 2.9\,dex and 3.2\,dex. We will adopt the mean value of
log\,$N_{0}$(Li)\,=\,3.1 as the cosmic ``initial'' lithium abundance.

\begin{figure}
\centering
\includegraphics[width=8.8cm]{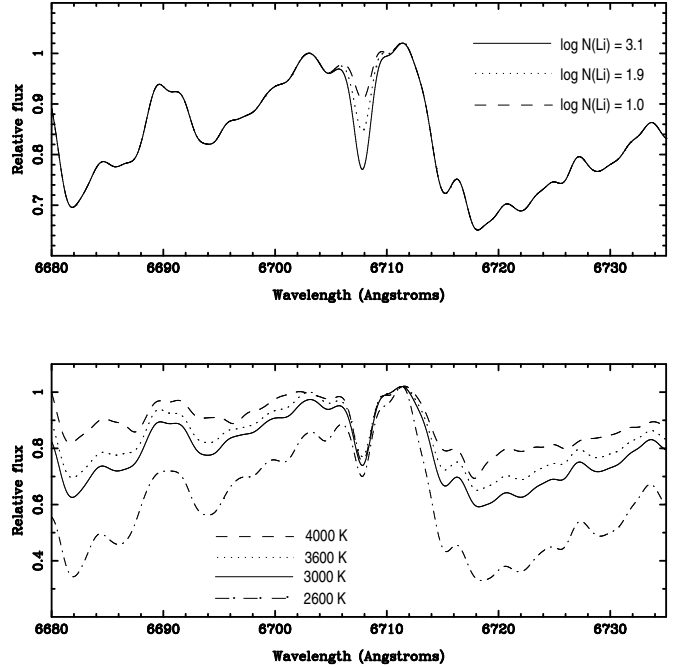}
\caption{\label{synspec} The upper panel shows theoretical 
  spectra computed for $T_{\rm eff}$\,=\,3400\,K, log\,$g$\,=\,4.0 and
  three different lithium abundances. The lower panel illustrates
  spectra for log\,$g$\,=\,4.0, log\,$N$(Li)\,=\,3.1 and various
  temperatures. The spectral resolution is $\sim$1.7\,\AA. The
  absorption feature centered at 6707.8\,\AA~is due to the atomic
  Li\,{\sc i} resonance doublet, while the rest of the spectral
  features are mainly molecular TiO absorptions.}
\end{figure}

\begin{table}
\caption[]{\label{theory} LTE Li\,{\sc i} 
$\lambda$6708\,\AA~resonance doublet curves of growth: predicted 
pseudo-equivalent widths (pEWs).}
\begin{tabular}{ccccccc}
\hline
\noalign{\smallskip}
\multicolumn{1}{c}{$T_{\rm eff}$ } &
\multicolumn{6}{c}{log\,$N$(Li)} \\
\multicolumn{1}{c}{(K)} &
\multicolumn{1}{c}{1.0} &
\multicolumn{1}{c}{1.6} &
\multicolumn{1}{c}{1.9} &
\multicolumn{1}{c}{2.5} &
\multicolumn{1}{c}{3.1} &
\multicolumn{1}{c}{3.4} \\
\noalign{\smallskip}
\hline
\noalign{\smallskip}
 2600 & .357 & .444 & .479 & .552 & .617/.644$^{\ast}$ & .656/.694$^{\ast}$ \\ 
 2800 & .346 & .440 & .475 & .551 & .623/.675$^{\ast}$ & .669/.728$^{\ast}$ \\ 
 3000 & .312 & .404 & .441 & .522 & .596/.666$^{\ast}$ & .652/.741$^{\ast}$ \\ 
 3200 & .296 & .386 & .423 & .504 & .578/.639$^{\ast}$ & .637/.729$^{\ast}$ \\ 
 3400 & .266 & .350 & .385 & .456 & .544/.634$^{\ast}$ & .604/.720$^{\ast}$ \\ 
 3600 & .262 & .337 & .373 & .442 & .536/.566$^{\ast}$ & .609/.665$^{\ast}$ \\ 
 4000 & .189 & .281 & .319 & .403 & .507/.537$^{\ast}$ & .588/.620$^{\ast}$ \\ 
\hline
\noalign{\smallskip}
\end{tabular}
\\
{\sc NOTES} ---  pEWs are given in \AA. In all computations we have 
used log\,$g$\,=\,4.0 and solar metallicity, except for the columns 
labelled with an asterisk, where we have used log\,$g$\,=\,4.5.
\end{table}

\begin{figure}
\centering
\includegraphics[width=8.8cm]{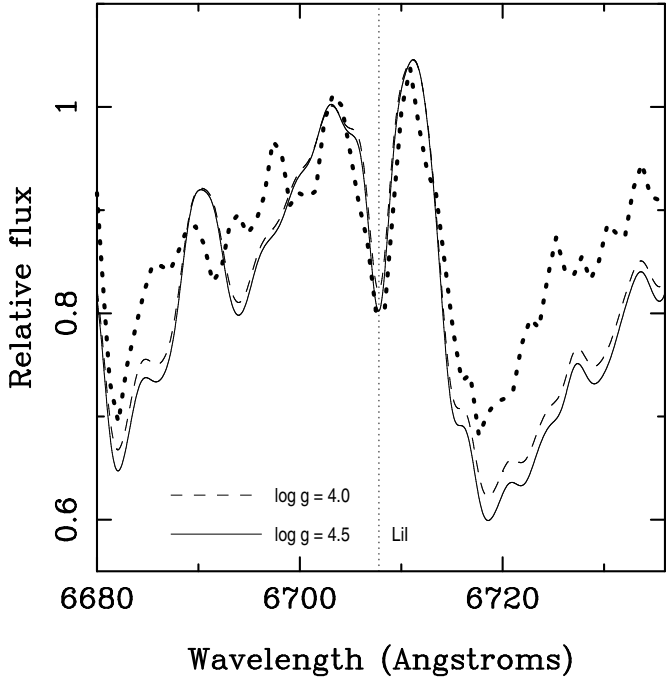}
\caption{\label{sori27} Synthetic spectra 
  ($T_{\rm eff}$\,=\,3000\,K, log\,$N$(Li)\,=\,3.1) compared to the
  observed spectrum of the brown dwarf S\,Ori\,27 (thick dotted line).
  Computed spectra have been degraded to the same resolution as the
  observations. The location of the Li\,{\sc i} resonance doublet is
  indicated with a vertical dotted line.}
\end{figure}

Figure~\ref{synspec} depicts some of our theoretical spectra for
different values of lithium abundance and surface temperature. The
observed spectrum of S\,Ori\,27 is compared to a few computations in
Fig.~\ref{sori27}. Optical spectra at these cool temperatures are
clearly dominated by molecular absorptions of TiO. Only the core of
the lithium line is observable, since the doublet wings are completely
engulfed by TiO lines (Pavlenko \cite{pav97}). We have obtained the
theoretical Li\,{\sc i} $\lambda$6708\,\AA~pEWs via direct integration
of the line profile over the spectral interval 6703.0--6710.8\,\AA.
Many of the lithium LTE curves of growth employed in this work are
presented in Table~\ref{theory}. Various authors (e.g., Magazz\`u,
Rebolo \& Pavlenko \cite{magazzu92}; Mart\'\i n et al$.$
\cite{martin94}; Pavlenko et al$.$ \cite{pav95}; Pavlenko
\cite{pav98}) have shown that the differences between LTE and non-LTE
calculations for cool temperatures are negligible compared to
uncertainties of pEW, $T_{\rm eff}$ and gravity. Similarly, the
effects of chromospheric activity on the line formation are found to
be of secondary importance (Pavlenko et al$.$ \cite{pav95}; Houdebine
\& Doyle \cite{houdebine95}; Pavlenko \cite{pav98}) and have not been
included in our calculations. The Li\,{\sc i} resonance doublet
appears to have very light dependence on the temperature structure of
the outer layers (see also Stuik, Bruls \& Rutten \cite{stuik97}). We
find a rather poor agreement between the predicted Li\,{\sc i} pEWs of
Table~\ref{theory} and those provided in Pavlenko \& Magazz\`u
(\cite{pav96}). These authors' values are considerably larger because
they measured theoretical equivalent widths (note the drop of
``pseudo'') relative to the computed ``real'' continuum, while we have
determined pEWs relative to the computed pseudo-continuum formed by
molecular absorptions.

\subsubsection{Observed spectra \label{lithium}}
We have also obtained the Li\,{\sc i} $\lambda$6708\,\AA~pEWs from our
observed spectra. To compensate for the different resolution of the
data, the integration of the line profile has always been performed
over the spectral range 6703.0--6710.8\,\AA. Our measurements and
their uncertainties are listed in Table~\ref{ew}.  Li\,{\sc i} is
detected in absorption in all of our program objects, except for
S\,Ori\,J053914.5--022834 (M3.5). It might be a cluster non-member,
but its optical spectrum is the noisiest amongst the McDonald data,
and even the Ca\,{\sc i} line at 6717\,\AA~lies barely undetected (see
Fig.~\ref{limcdonald}). We impose a 1\,$\sigma$ upper limit of
pEW\,=\,0.44\,\AA~by considering the strongest possible feature in the
region around the line and taking into account the S/N ratio and
resolution of the spectrum.

\begin{figure}
\centering
\includegraphics[width=8.8cm]{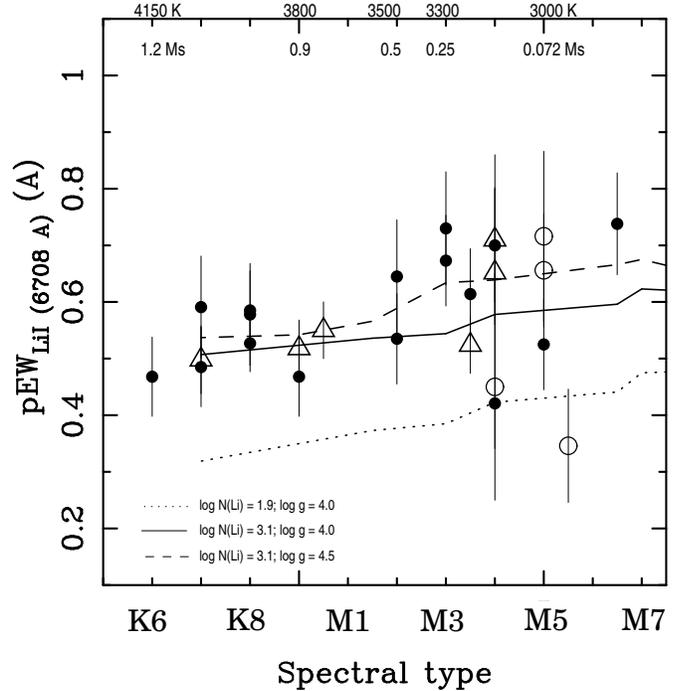}
\caption{\label{litsp} Pseudo-equivalent widths of Li\,{\sc i} 
  $\lambda$6708\,\AA~as a function of spectral type. Symbols are as in
  Fig.~\ref{ri}. Note that the coolest cluster member of our sample
  (the brown dwarf S\,Ori\,45) is not included in the figure for
  clarity. Overplotted onto the data are three LTE theoretical curves
  of growth provided in this paper. Typical uncertainty in spectral
  type is half a subclass. The stellar-substellar borderline takes
  place at M5--M6 spectral type at the age of the cluster. Effective
  temperatures in Kelvin and masses in solar units are also given.}
\end{figure}

Li\,{\sc i} pEWs are plotted against spectral type in
Fig.~\ref{litsp}. S\,Ori\,J053914.5--022834 is excluded from the
diagram. Overplotted onto the data are the theoretical pEWs for
log\,$g$\,=\,4.0 and two different lithium abundances:
log\,$N_{0}$(Li)\,=\,3.1 (``initial'') and log\,$N$(Li)\,=\,1.9 (about
one order of magnitude of destruction). We have also included in the
figure the ``initial'' curve of growth for a slightly larger gravity,
log\,$g$\,=\,4.5. The trend of the observations is nicely reproduced
by the log\,$N_{0}$(Li) curves, implying that lithium is still
preserved at the age of the $\sigma$\,Orionis cluster. We will discuss
this issue further in section~\ref{liage}. We note the differences due
to gravity in the Li\,{\sc i} curves of growth. Although these
differences are rather small (pEW\,$\le$\,0.03\,\AA) for $T_{\rm
  eff}$\,$\ge$\,3700\,K, they become twice as large for cooler
temperatures. Given the error bars of the observed Li\,{\sc i} pEWs,
we cannot easily discriminate between gravities.

The scatter of the Li\,{\sc i} pEWs is considerable for spectral types
cooler than M3.5 (Fig.~\ref{litsp}). The problem of the lithium
star-to-star dispersion occurring at $T_{\rm eff}$\,$\le$\,5300\,K has
been widely discussed in the literature (e.g., Soderblom et al$.$
\cite{soderblom93}; Pallavicini et al$.$ \cite{pallavicini93}; Russell
\cite{russell96}; Randich et al$.$ \cite{randich98}; Barrado y
Navascu\'es et al$.$ \cite{barrado01b}). Nevertheless, this phenomenon
still remains obscure and proves challenging to explain theoretically.
The dispersion could be ascribed to a variability in the Li\,{\sc i}
line as a consequence of stellar activity, different mixing processes,
presence or absence of circumstellar disks, binarity, or different
rotation rates from star to star. Recently, Fern\'andez \& Miranda
(\cite{fernandez98}) have found that the Li\,{\sc i}
$\lambda$6708\,\AA~line in the WTT star V410\,Tau varies according to
its rotational period. From Figs.~\ref{ha} and \ref{litsp} we observe
that the region of the largest lithium scatter coincides with that of
the strongest H$\alpha$ emissions. This might indicate that some hot
continuum is ``veiling'' the optical spectra (Joy \cite{joy45}; Basri
\& Batalha \cite{basri90}; Basri, Mart\'\i n \& Bertout
\cite{basri91}), thereby affecting our pEW measurements. We note,
however, that if any ``veiling'' exists around H$\alpha$ and Li\,{\sc
  i} in our spectra, it has to be small compared with that of many
other CTT stars, because there is no clear correlation between strong
H$\alpha$ emission and low values of Li\,{\sc i} pEWs (except for
S\,Ori\,J053951.6--022248). There are other possible explanations for
the significant Li\,{\sc i} pEW scatter, such as different gravities
(objects with low Li\,{\sc i} pEWs might have lower gravities, and
therefore, younger ages), and contamination by lithium-depleted
interlopers.


\section{Discussion \label{discussion}}
\subsection{Radial velocity: binarity}
\begin{figure}
\centering
\includegraphics[width=8.8cm]{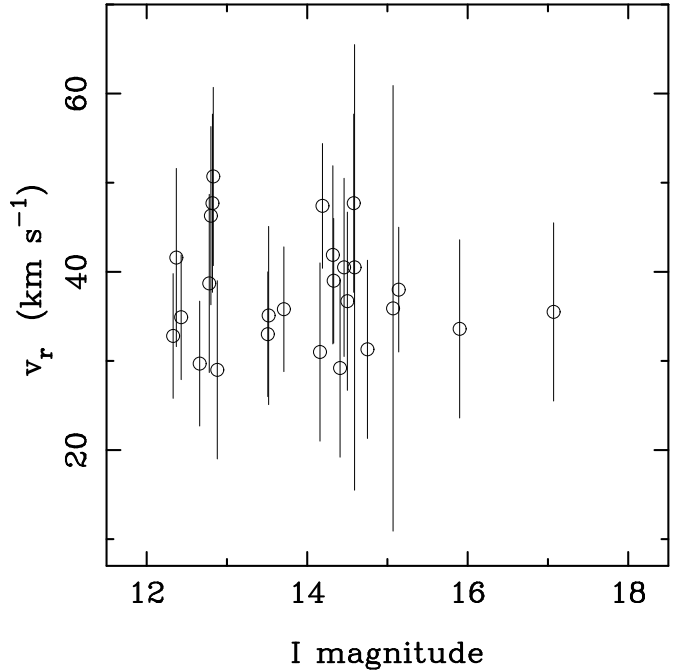}
\caption{\label{vrad} Radial velocities against $I$ magnitudes. 
  S\,Ori\,45 is not included in the figure (see text).}
\end{figure}

Lithium detections guarantee youth and the very likely membership of
our sample in the $\sigma$\,Orionis cluster. With radial velocities we
may be able to study possible multiplicity. However, the large
uncertainties and having only one epoch of observations for the
majority of the targets prevent us from carrying out a detailed
analysis. In general, the radial velocities in Table~\ref{ew} are in
the interval 30--50\,km\,s$^{-1}$. Walter et al$.$ (\cite{walter98})
obtained radial velocities of 104 pre-main sequence stars within
30\arcmin~from the $\sigma$\,Orionis star. These authors find a sharp
distribution peaking at around 25\,km\,s$^{-1}$ and covering a range
from 10\,km\,s$^{-1}$ up to 50\,km\,s$^{-1}$. Our measurements are in
full agreement with this wide radial velocity survey.
Figure~\ref{vrad} depicts our radial velocities against $I$
magnitudes. Neither drift nor an increasing dispersion are obvious at
the faintest magnitudes, indicating that the very low mass stars and
brown dwarfs of $\sigma$\,Orionis are not yet affected by internal
dynamical evolution, and that these objects still share the bulk
motion of the group.

Only the brown dwarf S\,Ori\,45 clearly shows a rather discrepant
radial velocity, which differs by more than 2.5\,$\sigma$ with respect
to the cluster mean velocity. With a mass estimated at around
0.02\,$M_{\odot}$ (B\'ejar et al$.$ \cite{bejar99}), S\,Ori\,45 is the
smallest object in our sample. It might belong to another kinematical
group of young stars, like the Taurus star-forming region or the Gould
Belt. On the basis of its multi-wavelength photometry and
spectroscopy, S\,Ori\,45 is probably not a member of Taurus. The
distance modulus to Taurus is 5.76 (Wichmann et al$.$
\cite{wichmann98}), which would make S\,Ori\,45 incredibly
overluminous by 2.2\,mag in the HR diagram.  Guillout et al$.$
(\cite{guillout98}) and Alcal\'a et al$.$ (\cite{alcala00}) have shown
that the distribution of candidate members of the Gould Belt for the
particular direction towards Orion lies at 200--300\,pc from the Sun
and well to the southwest of the Orion A cloud. This is relatively far
away from $\sigma$\,Orionis ($>$55\,pc). S\,Ori\,45 fits the
photometric and spectroscopic sequences of the $\sigma$\,Orionis
cluster very nicely (B\'ejar et al$.$ \cite{bejar99},
\cite{bejar01a}), supporting its location in the Orion complex.
Furthermore, this brown dwarf displays strong H$\alpha$ emission and
lithium in its atmosphere, which is typical of ages much younger than
that of the Gould Belt (30--80\,Myr, Alcal\'a et al$.$
\cite{alcala00}; Moreno, Alfaro \& Franco \cite{moreno99}), and it
does not show a radial velocity consistent with membership in either
Taurus or the Gould Belt. Alternatively, S\,Ori\,45 might be a runaway
object of the $\sigma$\,Orionis cluster resulting from encounters with
other cluster members; it may have been dynamically ejected from the
multiple system where it originated (Kroupa \cite{kroupa98}; Portegies
Zwart et al$.$ \cite{zwart99}; Reipurth \& Clarke \cite{reipurth01};
Boss \cite{boss01}), or S\,Ori\,45 might be a brown dwarf close
binary. So far none of these hypotheses can be discarded.  Further
radial velocity measurements are needed to assess the possible binary
nature. If S\,Ori\,45 is proved to be a spectroscopic binary, the
dynamical masses of the components will be valuable for testing
theoretical evolutionary tracks at very young ages and substellar
masses.

From Fig.~\ref{tspmag} we observe that r053820--0237 (M5) appears
remarkably overluminous with respect to the cluster photometric
sequence.  In addition, its radial velocity is the largest amongst our
measurements. These two properties suggest that this star is an equal
mass binary.

\subsection{The age of the $\sigma$\,Orionis cluster \label{liage}}
We do not observe from Fig.~\ref{litsp} that our $\sigma$\,Orionis
targets have undergone appreciable lithium destruction. Actually, the
Li\,{\sc i} curves of growth that neatly reproduce the observations
are those computed with the ``initial'' lithium abundance. More
massive F- and G-type stars in the Orion complex have similar lithium
contents (Cunha, Smith \& Lambert \cite{cunha95}). The lower envelope
to the distribution of the Li\,{\sc i} pEWs shown in Fig.~\ref{litsp}
could be described by the log\,$N$(Li)\,=\,1.9 curve of growth, i.e.,
lithium depleted by about one order of magnitude. We shall discuss the
likely age of the $\sigma$\,Orionis cluster on the basis of no lithium
destruction, and depletions by factors of 3 (logarithmic abundance of
$\sim$2.5\,dex) and 10 (logarithmic abundance of 2.0\,dex).

\begin{figure}
\centering
\includegraphics[width=8.8cm]{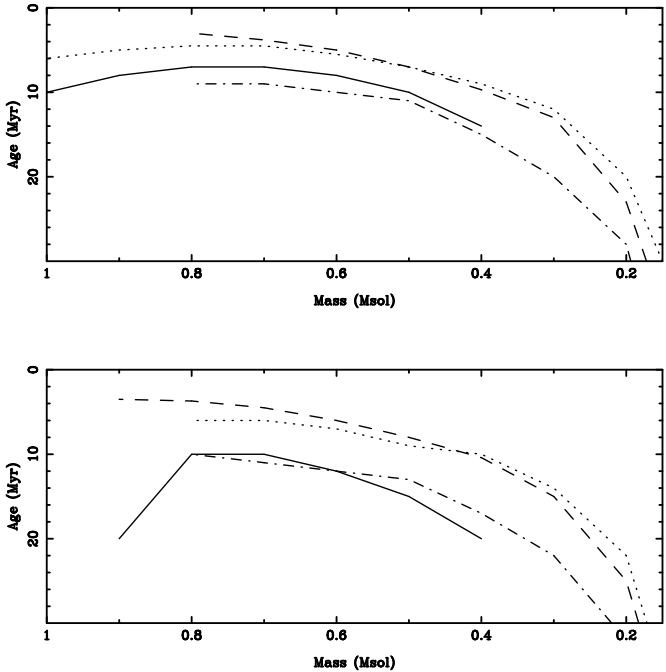}
\caption{\label{lidep} Surface curves for lithium depletions by 
  factors of 3 (log\,$N$(Li)\,=\,2.5, upper panel) and 10
  (log\,$N$(Li)\,=\,2.0, lower panel) as a function of age and
  mass. Models are taken from D'Antona \& Mazzitelli (\cite{dantona94},
  dashed line), D'Antona \& Mazzitelli (\cite{dantona97},
  dotted line), Pinsonneault et al$.$ (\cite{pin90}, solid line)
  and Baraffe et al$.$ (\cite{baraffe98}, dash-dotted line).}
\end{figure}

According to various evolutionary models available in the literature,
very low mass stars ($M$\,$\le$\,0.3\,$M_{\odot}$) burn lithium very
efficiently by one order of magnitude at ages older than 15\,Myr
(D'Antona \& Mazzitelli \cite{dantona94}, \cite{dantona97};
Pinsonneault et al$.$ \cite{pin90}; Baraffe et al$.$
\cite{baraffe98}). Stars with masses in the interval
0.5--0.8\,$M_{\odot}$ do it in a shorter time scale. This is
summarized in Fig.~\ref{lidep}, which shows surface curves for a given
lithium abundance as a function of age and stellar mass. The age of
the $\sigma$\,Orionis cluster will be constrained by late-K and
early-M stars. More massive members ($M$\,$\ge$\,0.9\,$M_{\odot}$)
need longer times to deplete some lithium, so they are not useful for
our purposes.

Lithium depletion by a factor of 10 will impose a rather conservative
upper limit on the age of the cluster. From Fig.~\ref{lidep} we infer
that this upper limit is around 10\,Myr (based on Baraffe et al$.$
\cite{baraffe98} and Pinsonneault et al$.$ \cite{pin90} models),
because this is the time required by 0.6--0.8\,$M_{\odot}$-stars to
consume their lithium from initial abundance down to
log\,$N$(Li)\,=\,2.0. Models by D'Antona \& Mazzitelli
(\cite{dantona94}, \cite{dantona97}) predict values that are twice as
young, i.e., around 5\,Myr. However, since no lithium depletion is
apparent in any cluster member, it seems reasonable to establish
shorter upper limits. If we adopt the surface curve corresponding to a
factor of 3 lithium depletion, the plausible oldest age of the
$\sigma$\,Orionis cluster is 8\,Myr (as given by Baraffe et al$.$
\cite{baraffe98} and Pinsonneault et al$.$ \cite{pin90} models). We
have also inspected the lithium depletion tracks provided by Proffitt
\& Michaud (\cite{proffitt89}) and Soderblom et al$.$
(\cite{soderblom98}) obtaining very similar values. Our result fully
agrees with the maximum age expected for the central, most massive
cluster star to blow up as a supernova (Meynet et al$.$
\cite{meynet94}). $\sigma$\,Orionis low mass stars span an age range
similar to that of the early-type members, i.e., the low and high mass
populations are essentially coeval. Similar upper limits are found for
other associations in Orion, like the star-forming region around the
$\lambda$\,Orionis star (7--8\,Myr, Mathieu, Dolan \& Robert
\cite{mathieu01}), and Orion WTT stars (Alcal\'a, Chavarr\'\i a-K., \&
Terranegra \cite{alcala98}). We could adopt as the mean cluster age
the oldest isochrone for which lithium is still preserved within
0.2\,dex across the entire mass range. This occurs at roughly
2--4\,Myr considering all models, a result in full consistency with
previous analysis of theoretical isochrone fitting to the observed
photometry (B\'ejar et al$.$ \cite{bejar99}).

An additional constraint to the age of the cluster comes from the
ratio of CTT stars to WTT stars. Based on strong H$\alpha$ emission
and the presence of forbidden emission lines, this ratio turns out to
be in the range 30--40\%~in $\sigma$\,Orionis. Follow-up observations
of our targets (mid-infrared, radio) are, however, desirable to
confirm the presence of circumstellar disks. The ratio obtained in
$\sigma$\,Orionis is slightly smaller than that of younger regions,
like the area around the Orion Molecular Cloud (ratio $\ge$40\%,
1--3\,Myr, Rebull et al$.$ \cite{rebull00}), and considerably larger
than the one of older clusters and associations, like the Sco-Cen OB
association (ratio of 11\%), whose population of CTT stars, WTT stars
and post-T\,Tauri stars has been investigated by Mart\'\i n
(\cite{martin98}). This author defines post -T\,Tauri stars as young,
late-type stars that are burning lithium and display moderate
H$\alpha$ emission. The average age of the whole Sco-Cen OB
association is in the range 5--15\,Myr, as determined by de Geus, de
Zeeuw \& Lub (\cite{geus89}). We do not find evidence for the
existence of post-T\,Tauri stars in $\sigma$\,Orionis, and hence, this
cluster is essentially younger than the Sco-Cen OB association.

\section{Summary and conclusions \label{conclusions}}
We have presented intermediate- and low-resolution optical spectra
between 6100\,\AA~and 7000\,\AA, covering H$\alpha$ and Li\,{\sc i} at
$\lambda$6708\,\AA, for a total of 25 low mass stars and 2 brown
dwarfs members of the $\sigma$\,Orionis young star cluster. Spectral
types have been derived and are found to be in the interval K6--M8.5,
which corresponds to masses from 1.2\,$M_{\odot}$ down to
0.02\,$M_{\odot}$ after comparison with state-of-the-art evolutionary
models (Baraffe et al$.$ \cite{baraffe98}; Chabrier et al$.$
\cite{chabrier00}). We have measured radial velocities and
pseudo-equivalent widths (pEWs) of the H$\alpha$ and Li\,{\sc i}
atomic lines and find that all our targets show remarkable H$\alpha$
emission and Li\,{\sc i} $\lambda$6708\,\AA~in absorption.  All radial
velocities (except for one object) are consistent with membership in
the $\sigma$\,Orionis cluster as well as in the Orion complex. The
distribution of H$\alpha$ and Li\,{\sc i} pEWs against spectral type
exhibits a large scatter at classes cooler than M3.5.  This phenomenon
occurs at the approximate mass, $\sim$0.25\,$M_{\odot}$, where low
mass stars are expected to become fully convective. Some of our
objects also show emissions of He\,{\sc i} and forbidden emission
lines of [O\,{\sc i}], [N\,{\sc ii}] and [S\,{\sc ii}], probably
indicating accretion from circumstellar disks.  We infer that the
likely rate of $\sigma$\,Orionis low mass members resembling classical
T\,Tauri stars is in the range 30--40\%, suggesting that the cluster
is only a few Myr old.

We note the intriguing case of the coolest object in our sample,
S\,Ori\,45, an M8.5-type brown dwarf with a mass estimated at
0.02\,$M_{\odot}$ (B\'ejar et al$.$ \cite{bejar99}). It has a very
intense, variable H$\alpha$ emission and lithium in absorption. Our
tentative detection of forbidden emission lines of [N\,{\sc ii}] and
[S\,{\sc ii}] suggests that S\,Ori\,45 may have a cool, surrounding
disk from which it is accreting. This brown dwarf also displays a
radial velocity that deviates significantly from the cluster mean
velocity.

We have also presented very recent computations of Li\,{\sc i}
$\lambda$6708\,\AA~curves of growth for low gravities (log\,$g$\,=4.0
and 4.5), cool temperatures ($T_{\rm eff}$ \,=\,4000--2600\,K), and
lithium abundances in the interval log\,$N$(Li)\,=\,1.0--3.4. The
distribution of our observed Li\,{\sc i} pEWs appears to be well
reproduced by the theoretical pEWs computed for the cosmic lithium
abundance of log\,$N_{0}$(Li)\,=\,3.1. This leads us to conclude that
lithium has not yet been depleted in the $\sigma$\,Orionis cluster.
Therefore, after comparison to various lithium depletion curves
available in the literature, we impose an upper limit to the cluster
age of 8\,Myr, while the most likely age is in the interval 2--4\,Myr.

\begin{acknowledgements}
  We are thankful to I. Baraffe and colleagues for making electronic
  files of their evolutionary models available to us, and to Louise
  Good for correcting the English language used in the manuscript.  We
  also thank the staff at McDonald Observatory, especially David Doss,
  for their helpful assistance.  This research has made use of the
  SIMBAD database, operated at CDS, Strasbourg, France. Partial
  financial support was provided by the Spanish DGES PB98-0531-C02-02.
  YP acknowledges partial financial support from Small Research Grant
  of American Astronomical Society.  CAP acknowledges partial
  financial support from NSF (AST-0086321).

\end{acknowledgements}

\end{document}